\documentclass[manuscript]{aastex631}

\shorttitle{Multiple Giant Impacts and Chemical Equilibria}
\shortauthors{Maeda \& Sasaki}

\begin{document}

\title{Multiple Giant Impacts and Chemical Equilibria: An Integrated Approach to Rocky Planet Formation}

\author[0009-0006-0867-8758]{Haruya Maeda}
\affiliation{Department of Astronomy, Kyoto University, Kitashirakawa-Oiwake-
cho, Sakyo-ku, Kyoto 606-8502, Japan}
\email{haruhi@kusastro.kyoto-u.ac.jp}

\author[0000-0003-1242-7290]{Takanori Sasaki}
\affiliation{Department of Astronomy, Kyoto University, Kitashirakawa-Oiwake-
cho, Sakyo-ku, Kyoto 606-8502, Japan}

\begin{abstract}
During the formation of rocky planets, the surface environments of growing protoplanets were dramatically different from those of present-day planets. The release of gravitational energy during accretion would have maintained a molten surface layer, forming a magma ocean. Simultaneously, sufficiently massive protoplanets could acquire hydrogen-rich proto-atmospheres by capturing gas from the protoplanetary disk. Chemical equilibration among the atmosphere, magma ocean, and iron core plays a key role in determining the planet's interior composition. In this study, we investigate terrestrial planet formation under such primitive surface conditions. We conduct $N$-body simulations to model the collisional growth from protoplanets to planets, coupled with chemical equilibrium calculations at each giant impact event, where surface melting occurs. Our results show that planetary growth proceeds through a series of giant impacts, and the timing of these impacts relative to the dissipation of disk gas significantly influences the volatile budget. In particular, initial impacts, occurring while nebular gas is still present, can lead to excess hydrogen incorporation into the protoplanet's core. Subsequent impacts with hydrogen-poor bodies, after gas dispersal, can dilute this hydrogen content. This process allows for the formation of a planet with a hydrogen inventory consistent with Earth's current core. Our findings suggest that late giant impacts, occurring after the depletion of nebular gas, provide a viable mechanism for producing Earth-like interior compositions near 1 AU.
\end{abstract}


\section{Introduction} \label{sec:intro}

In the classical planet formation theory, kilometer-sized planetesimals collide and merge to form protoplanets of approximately $0.1\, M_{\oplus}$ at 1 AU \citep{kokubo00}. These protoplanets develop high-temperature magma oceans as gravitational energy is released during giant impact events, establishing layered structures that include a primordial hydrogen-rich atmosphere, a gravity-stratified iron core, and a magma ocean. Such stratification facilitates diverse chemical reactions across layers and is hypothesized to shape the intricate terrestrial environment we observe today \citep{young23}. As the magma ocean cools and solidifies, the primordial hydrogen atmosphere transitions into a secondary atmosphere via degassing \citep{ozima93}, and water vapor in the hot atmosphere condenses and precipitates to form oceans \citep{matsui86}. Together, these processes offer a foundational framework for interpreting the chemical composition and evolution of terrestrial planets.

A key question in planet formation is how the bulk chemical characteristics of terrestrial planets, including volatile contents, are determined. Early studies demonstrated that forming planets likely contain sufficient water to reduce the necessity for substantial late veneers \citep{bond2010}. By focusing on metal–silicate equilibria at the core–mantle boundary, numerical simulations combined with high-temperature, high-pressure experiments have shown how parameters such as pressure, temperature, and oxygen fugacity influence the final distribution of elements between core and mantle \citep{fischer2017}. In addition, Earth's volatile element abundances appear to reflect the accretion history of solid materials, reproducing observed abundances in the bulk silicate Earth \citep{sakuraba2021}. These findings collectively underscore that planetary chemical histories can be traced from their earliest assembly stages.

The origin of Earth's water has been another central puzzle. Various mechanisms have been proposed, including the inward transport of water-bearing planetesimals and embryos from beyond the snow line \citep{raymond2004, Izidoro2022}. Meanwhile, an alternative source involves in situ water production through hydrogen-oxygen reactions \citep{ikoma06}, whereby hydrogen from the primordial atmosphere dissolves in the magma ocean and subsequently migrates into the iron core \citep{iizuka17}. In parallel, silicate minerals may oxidize metallic iron within the interior, elevating the planet's overall oxidation state \citep{javoy10}. These interconnected processes demonstrate that interactions among a planet's interior, surface, and atmosphere play a decisive role in setting its final chemical environment.

Importantly, terrestrial planet formation occurs in tandem with the evolution of the surrounding protoplanetary gas disk. Gas disk dissipation typically concludes on a timescale of a few million years \citep{haisch01}, and this dissipation affects not only where and when giant impacts can occur but also how protoplanets acquire, lose, or replenish their primordial atmospheres. Giant impacts strong enough to melt or vaporize planetary surfaces can drive off pre-existing atmospheres \citep{genda05}, after which the planet may capture new gas from the disk if it persists. Collisions also inject significant energy, potentially returning the planet to a magma ocean state, resetting or modifying the chemical pathways that determine water production, iron core density deficits, and oxygen fugacity. Since gas dissipation timescales and giant impact timescales can overlap \citep{chambers98}, the timing of these events is critical to understanding the final terrestrial planet properties.

\citet{kominami02, kominami2004} examined the behavior of the gas drag force and computed the orbital dynamics of protoplanets within a residual amount of disk gas, focusing on whether Earth-sized planets with low eccentricities could form after most of the disk gas had dissipated. They showed that the gas dissipation destabilizes the protoplanet's orbit and causes a giant impacts, but that the drag force of disk gas remains strong enough to dump the protoplanet's eccentricities even after the series of giant impacts. They argued that applying a constant gas surface density or a slowly decreasing one would merely shift the collision timing without dramatically altering the final outcome. Moreover, under the assumption that $v_{\mathrm{gas}} = v_{\mathrm{kep}}$, the inward spiral due to sub-Keplerian gas velocity was not explored in detail. 

As a similar issue to the one addressed by \citet{kominami02, kominami2004}, it is still unknown whether there is still enough gas left to build up the chemical composition of the Earth when the disk gas is dissipated to the extent that giant impacts begin. Recent work by \citet{young23} addressed the complex chemical reactions on proto-Earth by modeling a three-layer protoplanet—a hydrogen-rich atmosphere, a surface magma ocean, and an iron core—and computing chemical equilibria among these layers. Their results highlight how convection-driven mixing within the magma ocean governs the timescale of chemical equilibration ($\sim 10^4 \,\text{yr}$). Remarkably, their equilibrium abundances of water, core density deficits, and redox states show strong similarity to present-day Earth. These findings suggest that the interior, surface, and atmosphere of rocky planets are deeply coupled from an early stage, and that it is feasible to model such planets by parameterizing their masses, temperatures, and initial compositions.

However, \citet{young23} explored a single equilibrium state for a protoplanet of $0.5\, M_{\oplus}$ with a hydrogen-rich atmosphere of 0.2 wt\% and didn't consider the atmospheric acquisition process. In reality, the mass of the protoplanet and the fraction of captured disk gas can vary significantly, especially under the influence of giant impacts that occur while the gas disk is still present. Since hydrogen content in the primordial atmosphere directly influences water production and iron core density deficits, capturing the protoplanetary disk's time-evolving gas profiles becomes essential for understanding the range of possible outcomes.

Here, we extend the approach of \citet{young23} into a fully time-resolved context by integrating repeated chemical equilibrium calculations into an N-body simulation of protoplanet growth. We build upon the orbital evolution model of \citet{kominami02}, in which dissipating gas exerts drag forces on protoplanets, stabilizing orbits and regulating the timing of giant impacts. By coupling these orbital dynamics with geochemical equilibria at each impact event, we aim to reveal how the interplay between disk gas evolution, impact chronology, and chemical reactions shapes the final composition of Earth-like planets. In particular, we track the iron core density deficit as a key observable, because it directly reflects processes such as hydrogen ingress into the core. While water production is also critical, Earth's water supply likely has multiple contributors, and thus the amounts we estimate here serve as a lower limit.

In what follows, Section \ref{sec:method} describes our $N$-body simulation framework with disk model and the geochemical equilibrium calculations. We then present the results of our orbital simulations and discuss the corresponding chemical equilibria in Section \ref{sec:result}. Section \ref{sec:discussion} provides discussions of broader implications. Finally, Section \ref{sec:conclusion} summarizes our key findings and suggests future directions.

\section{Methods} \label{sec:method}

This study consists of two models: $N$-body simulations and chemical equilibrium calculations. In this section, we describe these models and methods.

\subsection{N-body simulation} \label{sec:nbody}

We investigated the temporal evolution following the formation of protoplanets using an $N$-body simulation. The $N$-body calculations were performed using the GPLUM $N$-body code \citep{iwasawa16, ishigaki21}\footnote{GPLUM is not a direct $N$-body integrator but employs the tree method, which efficiently handles the short-term evolution of many-body systems. While this approach provides a good balance between computational efficiency and accuracy for our current study, direct $N$-body integrators may be more suitable for simulating long-term evolution. Future validation using a direct $N$-body code could be valuable for further confirming our results.}. The basic equation is:
\begin{equation}
H = \sum_{i}^{N} \frac{p_{i}^{2}}{2m_{i}} - \sum_{i=1}^{N-1} \sum_{j=i+1}^{N} \frac{G m_{i} m_{j}}{r_{ij}},
\end{equation}
where $H$ is the Hamiltonian, $G$ is the gravitational constant, $p_{i}$ is the momentum of particle $i$, $m_{i}$ is its mass, and $r_{ij}$ is the distance between particles $i$ and $j$.

Previous numerical simulations based on the classical theory of planet formation have demonstrated that, within the rocky planetary region of the solar system, the mass of a protoplanet, $M_{\mathrm{proto}}$, is determined by \citep{kokubo00}
\begin{equation}
M_{\mathrm{proto}} \simeq 0.1 \left( \frac{b}{10 \, r_{\mathrm{Hill}}} \right)^{\frac{3}{2}} \left(\frac{\Sigma_{\mathrm{solid}}}{10\, \mathrm{g}\, \mathrm{cm}^{-2}}\right)^{\frac{3}{2}} \left(\frac{a}{1 \, \mathrm{AU}}\right)^{\frac{3}{4}} M_{\oplus},
\label{eq:radius_mass_relation}
\end{equation}
as a function of the distance from the central star, $a$, where $\Sigma_{\mathrm{solid}}$ is the surface density of solid material at 1 AU \citep{hayashi81}, and $M_{\oplus}$ is Earth's mass. The surface density profile of solid material is proportional to $a^{-3/2}$ \citep{hayashi81}. The isolation mass depends on the orbital separation between protoplanets, $b$. Protoplanets with masses of approximately $0.1\,M_{\oplus}$ tend to aggregate at intervals of about 10 mutual Hill radii,
\begin{equation}
r_{\mathrm{Hill}} = \left(\frac{2M_{\mathrm{proto}}}{3 M_{\odot}}\right)^{\frac{1}{3}} a,
\end{equation}
where $M_{\odot}$ is the mass of the Sun. We assume that collisions between protoplanets always result in mergers without producing debris. Planetesimals and pebbles in the disk are assumed to have already been removed, and only interactions between protoplanets and between protoplanets and disk gas are considered.

The gas surface density of the protoplanetary disk, $\Sigma_{\mathrm{gas}}$, is determined by the orbital radius distribution and is initially set to 0.01--1 times the minimum-mass solar nebula (MMSN) value \citep{hayashi81},
\begin{equation}
\Sigma^{\mathrm{min}}_{\mathrm{gas}} = 1700 \left(\frac{a}{1 \,\mathrm{AU}}\right)^{-3/2} \, \mathrm{g\, cm^{-2}},
\end{equation}
and gas density at mid-plane of the disk,
\begin{equation}
\rho^{\mathrm{min}}_{\mathrm{gas}} 
= 1.0 \times 10^{-9}
\left(\frac{a}{1 \,\mathrm{AU}}\right)^{-11/4}
\, \mathrm{g \, cm^{-3}}.
\end{equation}
Because the gas surface density at the time of protoplanet formation remains uncertain, we varied the initial gas surface density as a parameter. We performed calculations over this entire range (see Appendix for details), but the main results presented in this study focus on the case where $\Sigma_{\mathrm{gas}}$ is set to 0.01 times the MMSN value.

The gas surface density is then assumed to dissipate exponentially over time:
\begin{equation}
\Sigma_{\mathrm{gas}}(t) = \Sigma_{\mathrm{gas}}(0) \exp \left(- \frac{t}{\tau_{\mathrm{diss}}}\right),
\end{equation}
where the dissipation timescale, $\tau_{\mathrm{diss}}$, is set to $10^{6}$ years as a typical value, reflecting the observed dissipation of disk gas within a few million years \citep{haisch01}. We also conducted calculations over a range of $\tau_{\mathrm{diss}} = 10^5\text{--}10^6$ years (see Appendix for details), but the main results presented in this study focus on the case where $\tau_{\mathrm{diss}} = 10^{6}$ years.

A gas drag model \citet{kominami02} introduced and we adopt in this study is based on the gravitational gas drag formula, not the aerodynamic gas drag
formula. The drag force exerted on a protoplanet by the disk gas is proportional to both the relative velocity between the gas and the protoplanet and the gas surface density \citep{kominami02}. Unlike \citet{kominami02}, who assumed strictly Keplerian gas motion, we account for the pressure gradient that reduces the gas velocity below the Keplerian speed. Specifically,
\begin{equation}
\boldsymbol{v}_{\mathrm{gas}} = \left( 1 - \eta \right) \boldsymbol{v}_{\mathrm{kep}},
\end{equation}
where $\boldsymbol{v}_{\mathrm{kep}} = \sqrt{GM_{\odot}/a}$ and $\eta$ is a factor representing the pressure gradient:
\begin{equation}
\eta = \frac{1}{2} \left( \alpha + \beta \right) \left( \frac{c_{\mathrm{s}}}{v_{\mathrm{kep}}} \right)^{2} \simeq 1.8 \times 10^{-3} \left( \frac{a}{1 \, \mathrm{AU}} \right)^{\frac{1}{2}} \left( \frac{L}{L_{\odot}} \right)^{\frac{1}{4}},
\end{equation}
with $\rho_{\mathrm{gas}} \propto a^{-\alpha}$, $T \propto a^{-\beta}$, and $c_{\mathrm{s}}$ the sound velocity. The gas drag force, $\boldsymbol{f}_{\mathrm{gas \, drag}}$, is given by \citep{kominami02}:
\begin{equation}
\boldsymbol{f}_{\mathrm{gas \, drag}} \equiv - \frac{\boldsymbol{v} - \boldsymbol{v}_{\mathrm{gas}}}{\tau_{\mathrm{grav}}},
\label{eq:gas_drag}
\end{equation}
where
\begin{equation}
\tau_{\mathrm{grav}} = 2500 \left( \frac{\Sigma_{\mathrm{gas}}}{\Sigma_{\mathrm{gas}}^{\mathrm{min}}} \right)^{-1} \left( \frac{M_{\mathrm{proto}}}{0.2 \, M_{\oplus}} \right)^{-1} \left( \frac{a}{1 \, \mathrm{AU}} \right)^{2} \, \mathrm{yr}.
\end{equation}
Since $\Sigma_{\mathrm{gas}}$ decreases exponentially with time, the drag force also diminishes accordingly.

Investigating the timing of giant impacts among protoplanets through these parameter studies is essential and aligns with our aim of understanding planet formation from a continuous time-series perspective. While \citet{kominami02} focused on whether Earth-sized planets with low eccentricities would eventually form and how the presence of disk gas influenced this outcome, our interest lies in the environmental conditions experienced by protoplanets throughout their formation process.

\begin{figure}[ht!]
\plotone{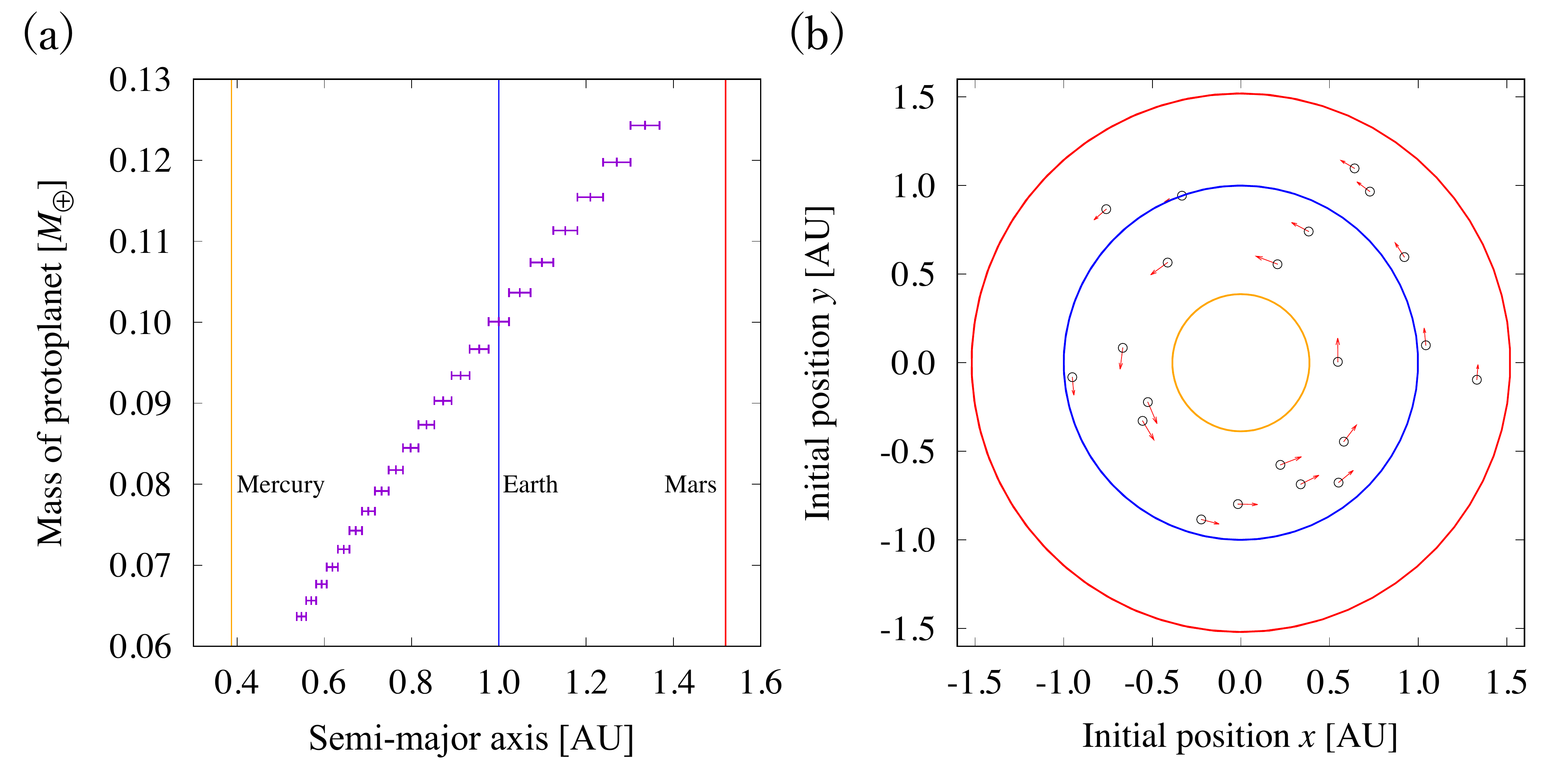}
\caption{(a) Initial semimajor axes and masses of protoplanets, and (b) their random distribution in the azimuthal direction, with arrows showing the velocity vectors. Yellow, blue, and red lines indicate the orbital radii of Mercury, Earth, and Mars, respectively. Error bars show the extent of solid material accumulation for each protoplanet, spanning $\pm 4\,r_{\mathrm{Hill}}$.
\label{fig:initial_orbit}}
\end{figure}

As initial conditions, \citet{kominami02} placed protoplanets with semimajor axis separations $\Delta a$ of $6\text{--}10\,r_{\mathrm{Hill}}$. Similarly, in this study, we start by placing a $0.1\,M_{\oplus}$ protoplanet at 1 AU. We then sequentially place additional protoplanets at 8\,$r_{\mathrm{Hill}}$ intervals on both sides of the initial one. This process continues until the total protoplanet mass reaches the combined mass of Mercury, Venus, Earth, and Mars ($M_{\mathrm{sum}} = 1.88 \, M_{\oplus}$). Consequently, the minimum and maximum orbital radii are 0.547 AU and 1.34 AU, respectively. We randomly distribute the 21 protoplanets azimuthally along their fixed orbital radii and assign each a Keplerian velocity at its position (Figure~\ref{fig:initial_orbit}).

\subsection{Geochemical equilibration} \label{sec:geochem}

In the chemical equilibrium model, we used the published code from \citet{young23} with some modifications. These chemical reactions are treated as optimization problems aimed at minimizing the Gibbs free energy to achieve chemical equilibrium. We solved the optimization problem using simulated annealing and Markov chain Monte Carlo (MCMC) methods. The basic equation is:

\begin{equation}
\sum_{i} \nu_{i} \log \nu_{i} + \left[ \frac{\Delta \hat{G}_{\mathrm{rxn}}^{0}}{RT} + \sum_{g}\nu_{g} \log\left( \frac{P}{P^{0}} \right) \right] = 0,
\end{equation}
where $\nu_{i}$ represents the stoichiometric coefficient for species $i$ in each reaction, $\Delta \hat{G}_{\mathrm{rxn}}^{0}$ is the molar Gibbs free energy, $P$ is the total gas pressure, $R$ is the gas constant, and $P^{0}$ is the standard-state pressure. The subscript $g$ refers to atmospheric species, while the subscript $i$ includes all species, both gaseous and non-gaseous. Table~\ref{tab:chemical_reaction} lists all reactions expected to occur on the protoplanet. 

The resulting equilibrium state is determined by the combination of equilibrium constants for each reaction across the parameter space. As a result, the internal structure of the model is somewhat obscured, functioning like a black box. In this sense, we focus on identifying which reactions are important for reproducing the present-day Earth, rather than on determining which reactions are dominant.

\begin{table}[h]
\begin{center}
\caption{Chemical reactions occurring on a protoplanet}
\begin{tabular}{l|c} \hline
   Location of reaction & Reaction \\
   \hline \hline
   Atmosphere & $2\mathrm{H_{2, gas}} + \mathrm{O_{2, gas}} \,\rightleftharpoons\, 2\mathrm{H_{2}O_{gas}}$ \\
   & $2\mathrm{CO_{gas}} + \mathrm{O_{2, gas}} \,\rightleftharpoons\, 2\mathrm{CO_{2, gas}}$ \\
   & $2\mathrm{CH_{4, gas}} + \mathrm{O_{2, gas}} \,\rightleftharpoons\, 4\mathrm{H_{2, gas}} + 2\mathrm{CO_{gas}}$ \\
   \hline
   Atmosphere $\leftrightarrow$ Magma Ocean & $\mathrm{H_{2, gas}} \,\rightleftharpoons\, \mathrm{H_{2, silicate}}$ \\
   & $\mathrm{H_{2}O_{gas}} \,\rightleftharpoons\, \mathrm{H_{2}O_{silicate}}$ \\
   & $\mathrm{CO_{gas}} \,\rightleftharpoons\, \mathrm{CO_{silicate}}$ \\
   & $\mathrm{CO_{2, gas}} \,\rightleftharpoons\, \mathrm{CO_{2, silicate}}$ \\
   & $2\mathrm{FeO} \,\rightleftharpoons\, 2\mathrm{Fe_{gas}} + \mathrm{O_{2, gas}}$ \\
   & $2\mathrm{MgO} \,\rightleftharpoons\, 2\mathrm{Mg_{gas}} + \mathrm{O_{2, gas}}$ \\
   & $2\mathrm{SiO} \,\rightleftharpoons\, 2\mathrm{SiO_{gas}} + \mathrm{O_{2, gas}}$ \\
   & $2\mathrm{Na_{2}O} \,\rightleftharpoons\, 4\mathrm{Na_{gas}} + \mathrm{O_{2, gas}}$ \\
   \hline
   Magma Ocean \& CMB & $\mathrm{Na_{2}SiO_{3}} \,\rightleftharpoons\, \mathrm{Na_{2}O} + \mathrm{SiO_{2}}$ \\
   & $\mathrm{FeSiO_{3}} \,\rightleftharpoons\, \mathrm{FeO} + \mathrm{SiO_{2}}$ \\
   & $\mathrm{MgSiO_{3}} \,\rightleftharpoons\, \mathrm{MgO} + \mathrm{SiO_{2}}$ \\
   & $\mathrm{SiO_{2}} + 2\mathrm{Fe_{metal}} \,\rightleftharpoons\, 2\mathrm{FeO} + \mathrm{Si_{metal}}$ \\
   & $2\mathrm{H_{2}O_{silicate}} + \mathrm{Si_{metal}} \,\rightleftharpoons\, \mathrm{SiO_{2}} + 2\mathrm{H_{2, silicate}}$ \\
   & $2\mathrm{H_{metal}} \,\rightleftharpoons\, \mathrm{H_{2, silicate}}$ \\
   & $2\mathrm{O_{metal}} + \mathrm{Si_{metal}} \,\rightleftharpoons\, \mathrm{SiO_{2}}$ \\
   \hline
\end{tabular}
\label{tab:chemical_reaction}
\end{center}
\tablecomments{Chemical reactions occur in the interior, on the surface, and within the atmospheres of protoplanets. Chemical species without subscripts refer to those in silicate.}
\end{table}

\begin{figure}[ht!]
\plotone{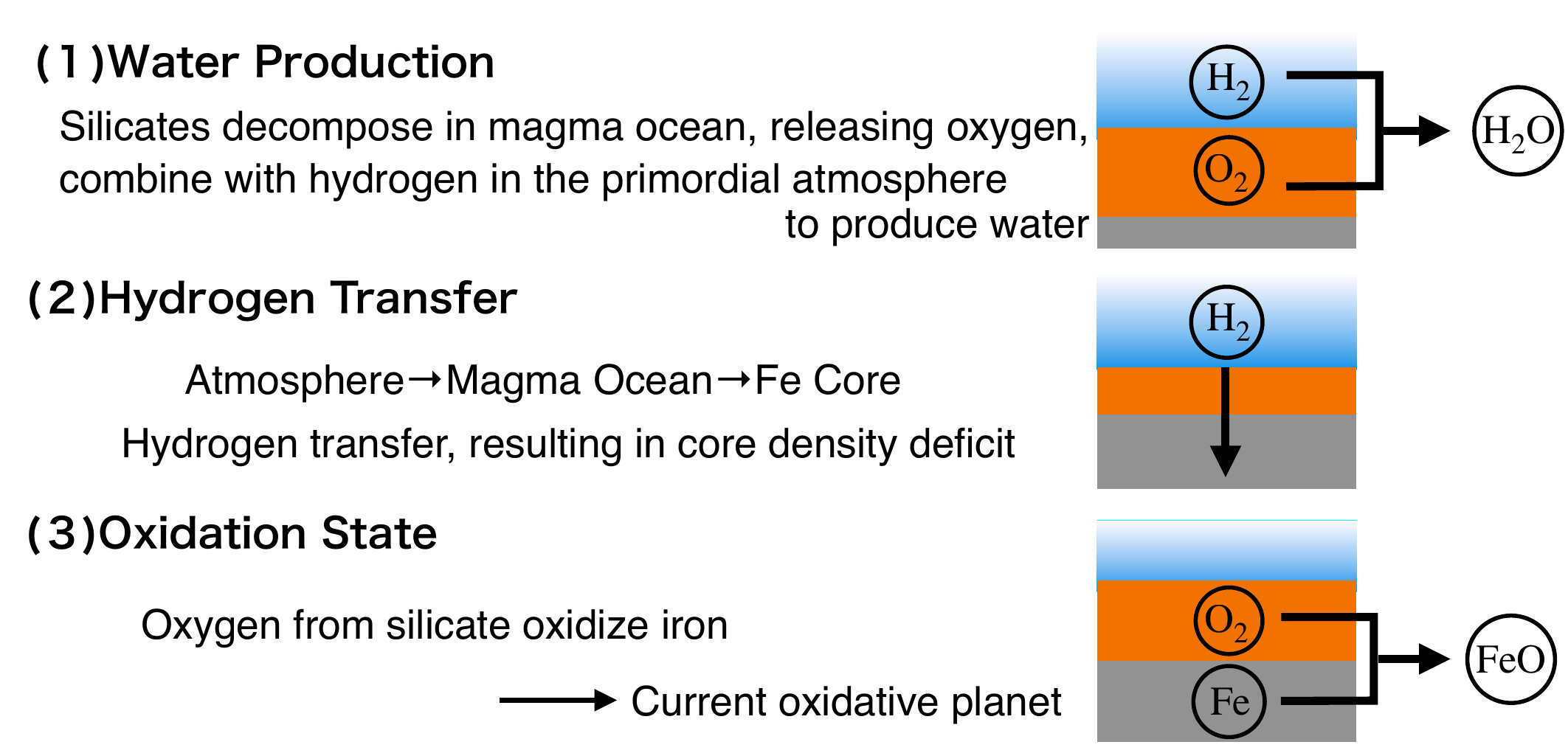}
\caption{Key reactions that produce Earth's current environment. (1) Silicates decompose in the magma ocean and release oxygen, which combines with atmospheric hydrogen to form water. (2) Hydrogen migrates from the atmosphere through the mantle into the core, decreasing the core density. (3) Silicate-derived oxygen oxidizes iron in the core, creating an overall more oxidized planet.
\label{fig:key_reactions}}
\end{figure}

The following three chemical reactions, occurring within each layer and at their boundaries, are most relevant for reproducing Earth's present composition (Figure~\ref{fig:key_reactions}):

(1) Water production: Silicate minerals decompose in the magma ocean, forming simpler oxides (MgO, FeO, Na$_2$O, SiO$_2$). These oxides can further decompose and outgas into the atmosphere. Once in contact with atmospheric hydrogen, water vapor is generated \citep{ikoma06}. Although our calculations do not simulate post-equilibrium cooling, this vapor would eventually condense and precipitate, ultimately forming oceans \citep{matsui86}.

(2) Hydrogen transfer: Atmospheric hydrogen dissolves into the magma ocean under high pressure and is subsequently incorporated into the iron core, resulting in a density deficit compared with pure iron. This is consistent with the density deficit in Earth's core \citep{birch1964}.

(3) Oxidation tagetate: As silicates in the magma ocean decompose, SiO$_2$ forms. Iron in the core partially reduces SiO$_2$, creating FeO in the magma ocean. Iron oxide then participates in additional reactions—some releasing metallic gas at high temperatures—leading to an FeO fraction that remains in the mantle as iron-bearing silicate minerals. Earth's composition is thus more oxidized than that of its bulk building materials \citep{javoy10}.

We assume that giant impacts release sufficient energy to completely melt the mantle, forming a global magma ocean. The protoplanet also acquires a hydrogen-rich atmosphere from the surrounding disk gas. Chemical equilibrium calculations are carried out for a protoplanet composed of three layers: atmosphere, magma ocean, and core, following \citet{young23} for the initial core and mantle compositions. The mass fraction of Fe metal in the protoplanet is set to 34.4 wt\%, reproducing the present-day Earth's core fraction (including Fe, Si, O, and H). The mantle composition reflects the bulk silicate Earth (BSE), consisting primarily of Mg, Si, Fe, Na, and O.

For the atmosphere, we assume that protoplanets capture all disk gas within a distance of $2\,r_{\mathrm{Hill, \,s}}$ (i.e., $\pm1\,r_{\mathrm{Hill, \,s}}$) based on the gas surface density at the time of a giant impact. Here, we use single Hill radius $r_{\mathrm{Hill, \,s}} \equiv (M_{\mathrm{proto}}/3M_{\odot})^{1/3}a$ because mass of gas molecule is sufficiently smaller than mass of protoplanet. The atmospheric mass, $M_{\mathrm{atm}}$, is given by
\begin{equation}
M_{\mathrm{atm}} = M_{\mathrm{torus}} 
= 2 \pi r \, \left(\pi r_{\mathrm{Hill, \,s}}^{2}\right)
\,\rho_{\mathrm{gas}},
\label{eq:Atm torus}
\end{equation}
where $r$ is the semimajor axis of the protoplanet and $\rho_{\mathrm{gas}}$ is the local gas density at the time of the giant impact. The local gas density is also assumed to dissipate exponentially over time: 
\begin{equation}
\rho_{\mathrm{gas}}(t) = \rho_{\mathrm{gas}}(0) \exp \left(- \frac{t}{\tau_{\mathrm{diss}}}\right).
\end{equation}

Under this formulation, the protoplanet captures disk gas in a toroidal volume. For example, for a protoplanet with mass $0.5\,M_{\oplus}$ at 1 AU, where the disk scale height greatly exceeds the Hill radius, the gas density $\rho_{\mathrm{gas}}$ can be treated as nearly uniform for the vertical direction inside the capture region. 

This assumption implies that gas within the Hill sphere can participate in chemical reactions, even if it is not permanently retained as a stable atmosphere, because hydrogen can be drawn into the interior while the protoplanet continues to accrete disk gas. This $M_{\mathrm{torus}}$ value gives a lower limit on the atmospheric mass under our assumptions using the amount of disk gas in the region through which the Hill sphere passes. However, gas inflow to a protoplanet occurs only at certain altitudes and has a complex three-dimensional structure \citep[e.g.,][]{Fung2015, Kuwahara2019}. Taking into account both the three-dimensional motion of the protoplanet and the altitude-dependent inflow, the situation becomes more complex than assumed here.

Here, we set another atmospheric mass value based on the gas surface density. When we add up all the atmospheric volume in the vertical direction, We get

\begin{equation}
M_{\mathrm{atm}} = M_{\mathrm{cylinder}} = 2 \pi r \Delta r \, \Sigma_{\mathrm{gas}} \quad (\Delta r = 2 \, r_{\mathrm{Hill, \,s}}),
\label{eq:Atm cylinder}
\end{equation}

where $\Sigma_{\mathrm{gas}}$ is the local gas surface density at the time of the giant impact. This value gives a upper limit on the atmospheric mass. In this study, we use three values for atmospheric mass in our calculations; $M_{\mathrm{torus}}$, $M_{\mathrm{cylinder}}$, and $M_{\mathrm{mid}} \equiv (M_{\mathrm{torus}} + M_{\mathrm{cylinder}})/2$ in between.

In reality, not all neighboring disk gas may be gravitationally retained. However, hydrogen consumed by equilibrium processes in the interior can be replenished from the disk on a timescale comparable to the $\sim10^4$-year convection timescale in the magma ocean for protoplanets with masses of 0.1 $M_{\oplus}$ and atmospheres of about 1 wt\%, which are typical in this study \citep{Ormel2015}. Thus, this scenario reasonably assumes that an atmosphere can be replenished at about the same speed relative to the rate-limiting steps of interior equilibration.

Although the actual process of gas accretion can be slower than mantle solidification \citep{ikoma06}, and 3D simulations show ``atmospheric recycling'' in which gas enters the Bondi sphere but later exits \citep[e.g.,][]{Ormel2015, Fung2015}, hydrogen that reaches the surface can still react and enter the interior on short timescales, potentially depleting the atmosphere and allowing more disk gas to replenish it. Thus, while the effective capture radius may be smaller than the Hill or Bondi radius, the net effect can still result in substantial hydrogen incorporation. We therefore adopt the Hill radius as the effective region within which disk gas can react with the protoplanet.

If the post-impact mass of a protoplanet is below $0.2\,M_{\oplus}$, we omit chemical equilibrium calculations, following \citet{young23}, because a body with a hot surface below this threshold is unlikely to retain a significant atmosphere. Although this cutoff is simplified, it serves as a practical modeling choice. Even if smaller protoplanets had partial atmospheres before the giant impact stage, the key factor here is that large impacts melt the lower mantle and deliver hydrogen to the core, making prior atmospheric states less critical.

Although several studies have investigated the detailed behavior of disk gas, the primary focus of this study is not merely whether the protoplanet binds and retains an atmosphere, but rather how much disk gas it can consume through chemical reactions.

\begin{figure}[ht!]
\plotone{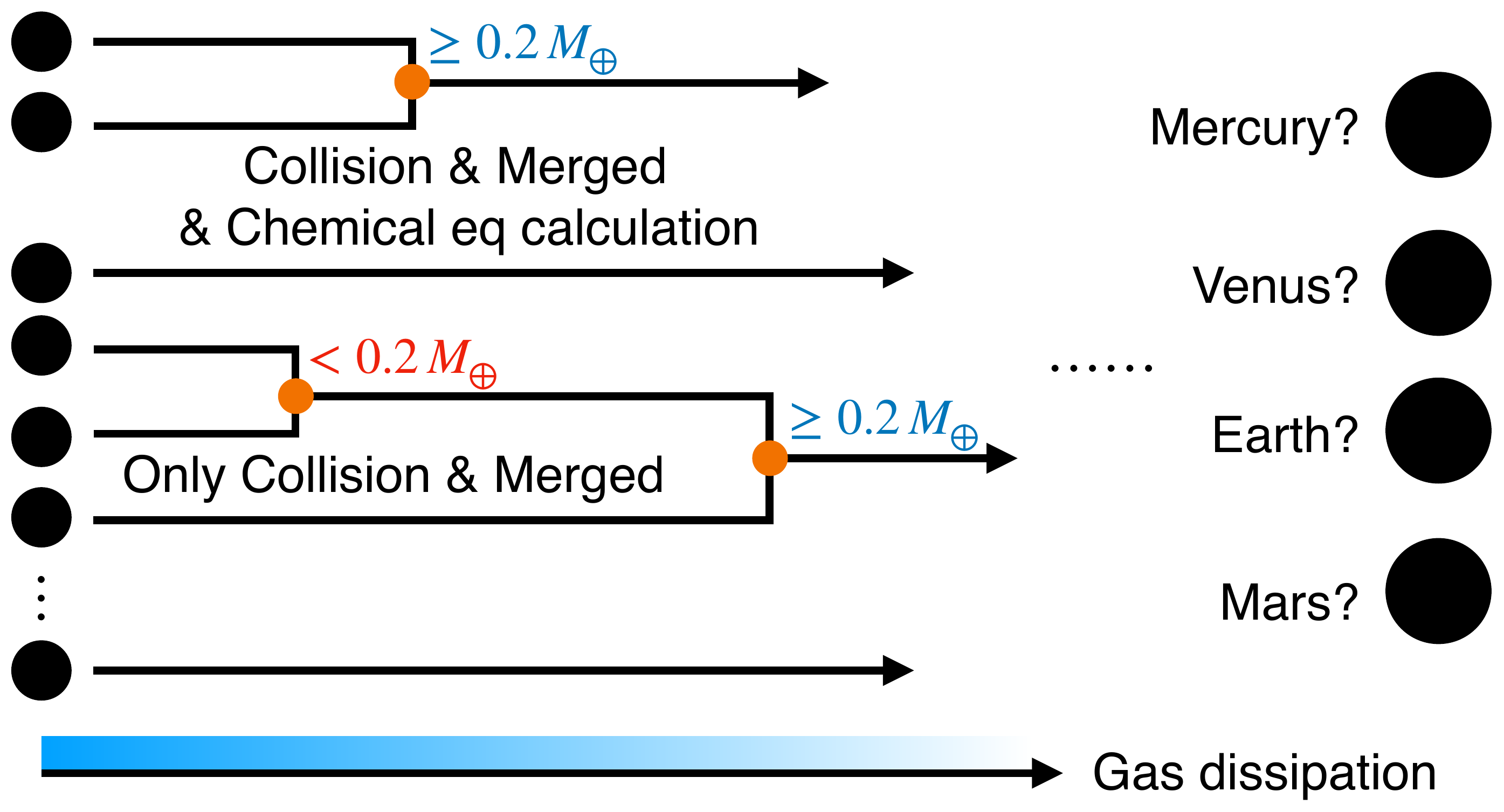}
\caption{Process flow diagram of our calculations. Initially, 21 protoplanets collide and merge due to mutual gravitational interactions, ultimately forming four rocky planets analogous to Mercury, Venus, Earth, and Mars. Meanwhile, the protoplanetary disk gas dissipates exponentially with time. Depending on disk conditions and the mass threshold for atmospheric capture, giant impacts repeatedly melt the mantle, triggering chemical equilibria. We perform equilibrium calculations at each giant impact event to update the protoplanet's composition accordingly.
\label{fig:process_calculation}}
\end{figure}

Because protoplanets typically undergo multiple giant impacts, we perform chemical equilibrium calculations sequentially for each impact event. The equilibrium result after one collision becomes the initial composition for the next (Figure~\ref{fig:process_calculation}). Giant impacts can also strip away atmospheric hydrogen, which is subsequently replenished by the surrounding disk gas.

\section{Results}\label{sec:result}

\subsection{Results of $N$-body simulation} \label{sec:result_nbody}

We first performed $N$-body calculations for a configuration with protoplanets initially placed with inclination $i=0$. The results of these calculations are summarized in the Appendix. We concluded that the overall system behavior does not change significantly across a wide range of disk parameters, except in extreme cases. Therefore, we adopted representative gas disk profiles for the subsequent calculations: an initial gas surface density of $\Sigma_{\mathrm{gas}}(0) = 0.01 \, \Sigma^{\mathrm{min}}_{\mathrm{gas}}$ and a gas dissipation timescale of $\tau_{\mathrm{diss}} = 10^{6}$ yr.

We randomly assigned small initial eccentricities ($e$) and inclinations ($i$) below 0.01, following \citet{kominami02}, to the semimajor axis and mass relations based on Eq.~\ref{eq:radius_mass_relation}. Under these conditions, we generated 12 sets of initial conditions and simulated them for $10^{8}$ years.

\begin{table}[h]
\begin{center}
\caption{$N$-body calculation results}
\begin{tabular*}{170mm}{@{\extracolsep{\fill}}ccccc} \hline
   Run & $N_{\mathrm{proto}}$ at $t = 1.6 \times 10^{6}$ yr & $N_{\mathrm{proto}}$ at $t = 3.2 \times 10^{6}$ yr & $N_{\mathrm{proto}}$ at $t = 1.6 \times 10^{7}$ yr & Final $N_{\mathrm{proto}}$\\
   \hline \hline
   No.1 & 12 (3) & 9 (5) & 7 (4) & 4 (3) \\
   No.2 & 11 (4) & 10 (3) & 7 (3) & 4 (3) \\
   No.3 & 13 (3) & 10 (3) & 7 (4) & 5 (3) \\
   No.4 & 11 (5) & 10 (5) & 9 (5) & 9 (5) \\
   No.5 & 10 (4) & 9 (5) & 9 (5) & 9 (5) \\
   \underline{No.6} & 10 (3) & 10 (3) & 6 (4) & 4 (3) \\
   \underline{No.7} & 10 (5) & 7 (5) & 6 (6) & 6 (6) \\
   No.8 & 10 (5) & 8 (5) & 6 (5) & 6 (5) \\
   No.9 & 9 (5) & 9 (5) & 9 (5) & 9 (5) \\
   No.10 & 11 (3) & 9 (3) & 6 (4) & 6 (4) \\
   \underline{No.11} & 12 (5) & 10 (5) & 6 (5) & 3 (2) \\
   \underline{No.12} & 13 (2) & 10 (3) & 8 (3) & 4 (3) \\
   \hline
\end{tabular*}
\label{tab:inclination N-body}
\end{center}
\tablecomments{For each run, the table shows the number of protoplanets at $t = 1.6 \times 10^{6}$ yr, $3.2 \times 10^{6}$ yr, $1.6 \times 10^{7}$ yr, and at $t_{\mathrm{end}} (10^{8} \,\mathrm{yr})$. Numbers in parentheses represent the number of protoplanets with a mass of $0.2 \, M_{\oplus}$ or more. The mean final number of planets is $\langle N_{\mathrm{planet}} \rangle = 5.75$. For the underlined cases (No.6, No.7, No.11, and No.12), the time evolution of orbital elements is shown in Figure~\ref{fig:orbital evolution for i not 0}.}
\end{table}

The results are listed in Table~\ref{tab:inclination N-body}. Although in some cases the final number of protoplanets was reduced to four, averaging over the 12 simulations, the final number of protoplanets ($\langle N_{\mathrm{planet}} \rangle = 5.75$) was larger than in calculations without gas drag \citep{kokubo06} (see Section~\ref{sec:final_number_planet}).

\begin{figure}[ht!]
\centering
\includegraphics[width=0.7\textwidth]{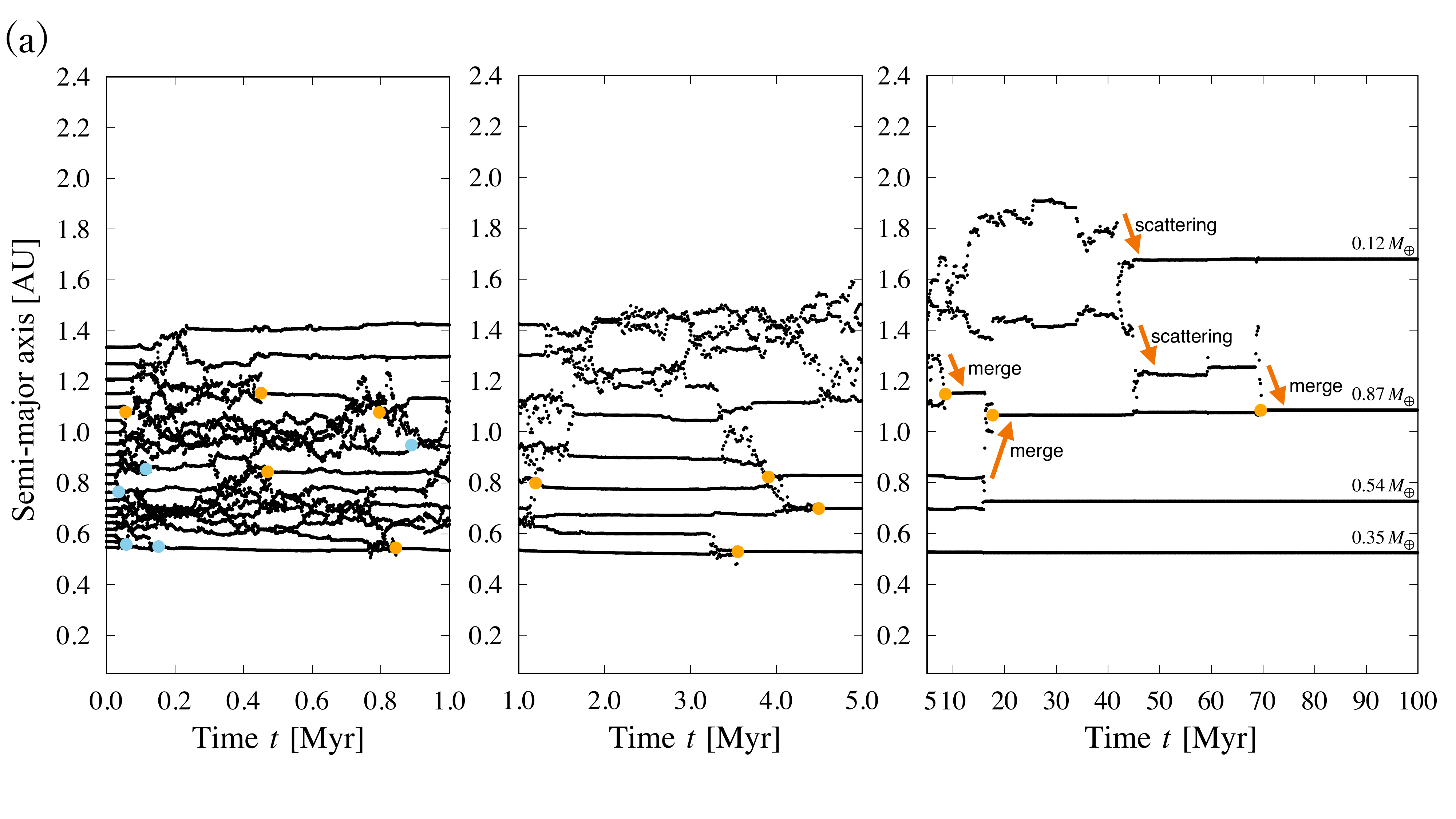}
\includegraphics[width=0.7\textwidth]{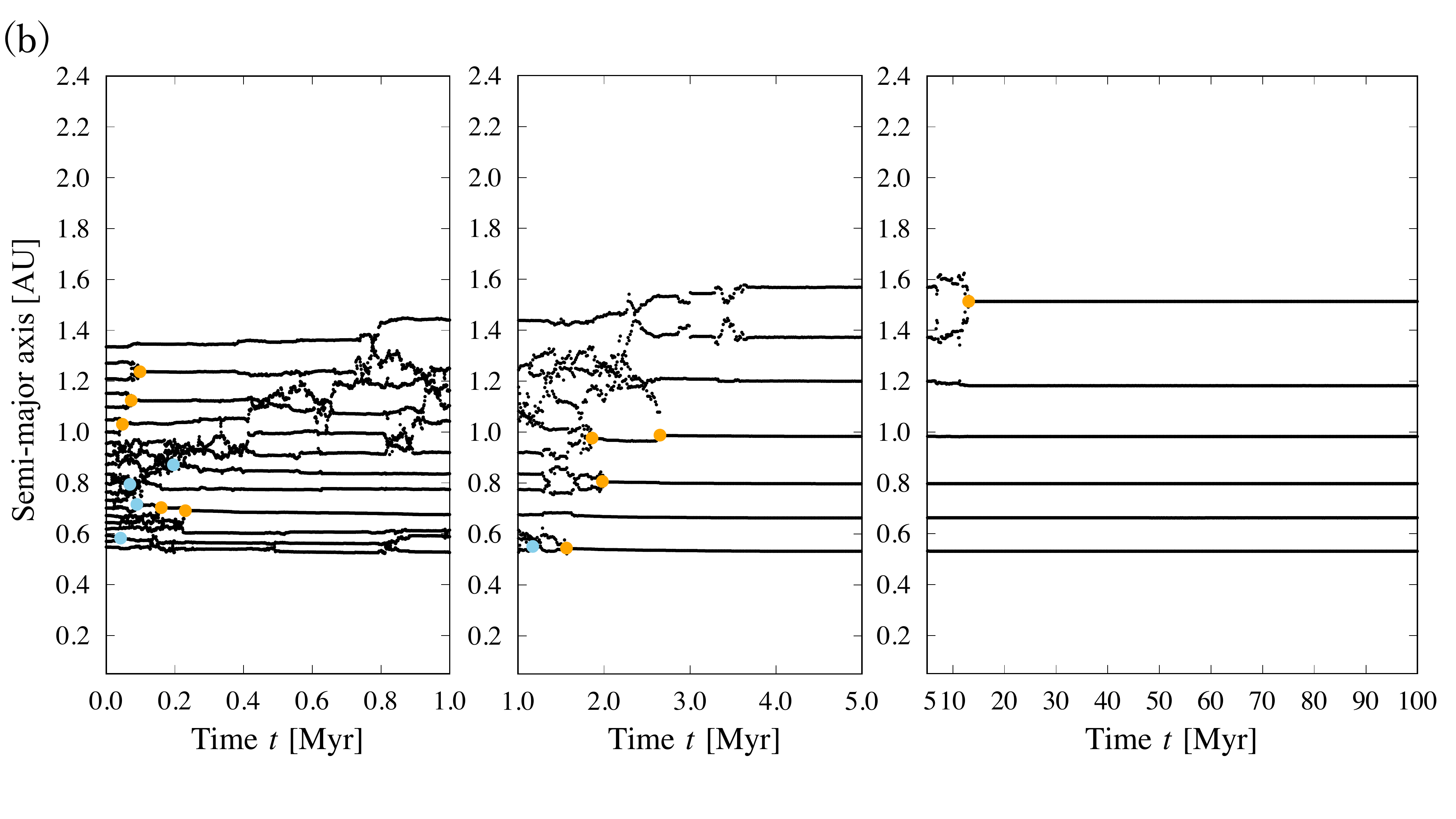}
\caption{Orbital evolution of protoplanets over the full simulation time. Colored circles indicate protoplanets immediately after experiencing giant impacts and mergers. Blue circles correspond protoplanets with mass less than $0.2 \, M_{\oplus}$ and orange ones correspond protoplanets with mass of $0.2 \, M_{\oplus}$ or more. (a) Case No.6, (b) Case No.7, (c) Case No.11, and (d) Case No.12 in Table~\ref{tab:inclination N-body}. The system rapidly becomes unstable, triggering initial giant impacts. Although about half of the giant impacts occur within the first million years, additional impacts continue to occur later, even beyond $10^{7}$ years.}
\label{fig:orbital evolution for i not 0}
\end{figure}

\begin{figure}[ht!]
\centering
\includegraphics[width=0.7\textwidth]{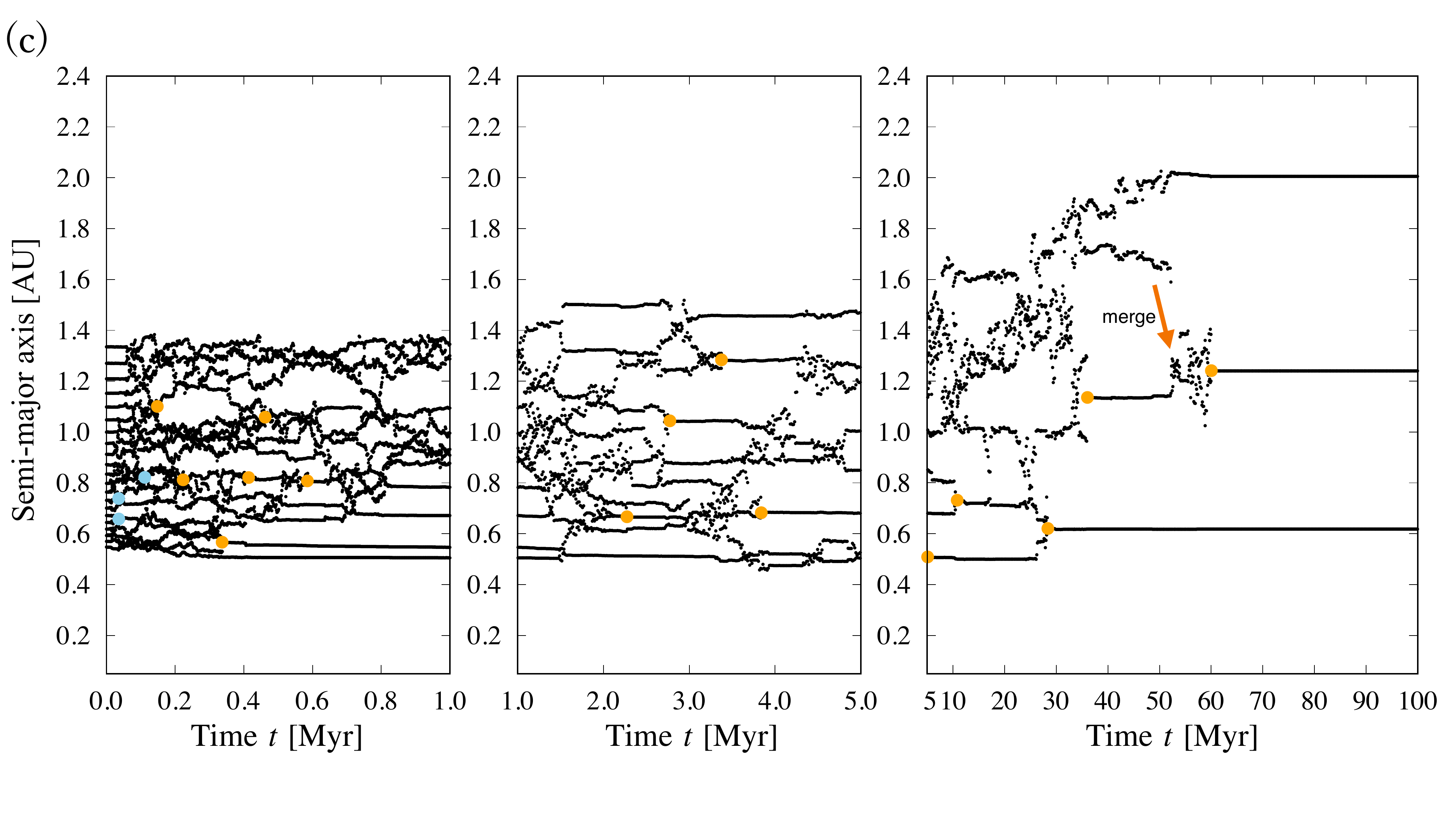}
\includegraphics[width=0.7\textwidth]{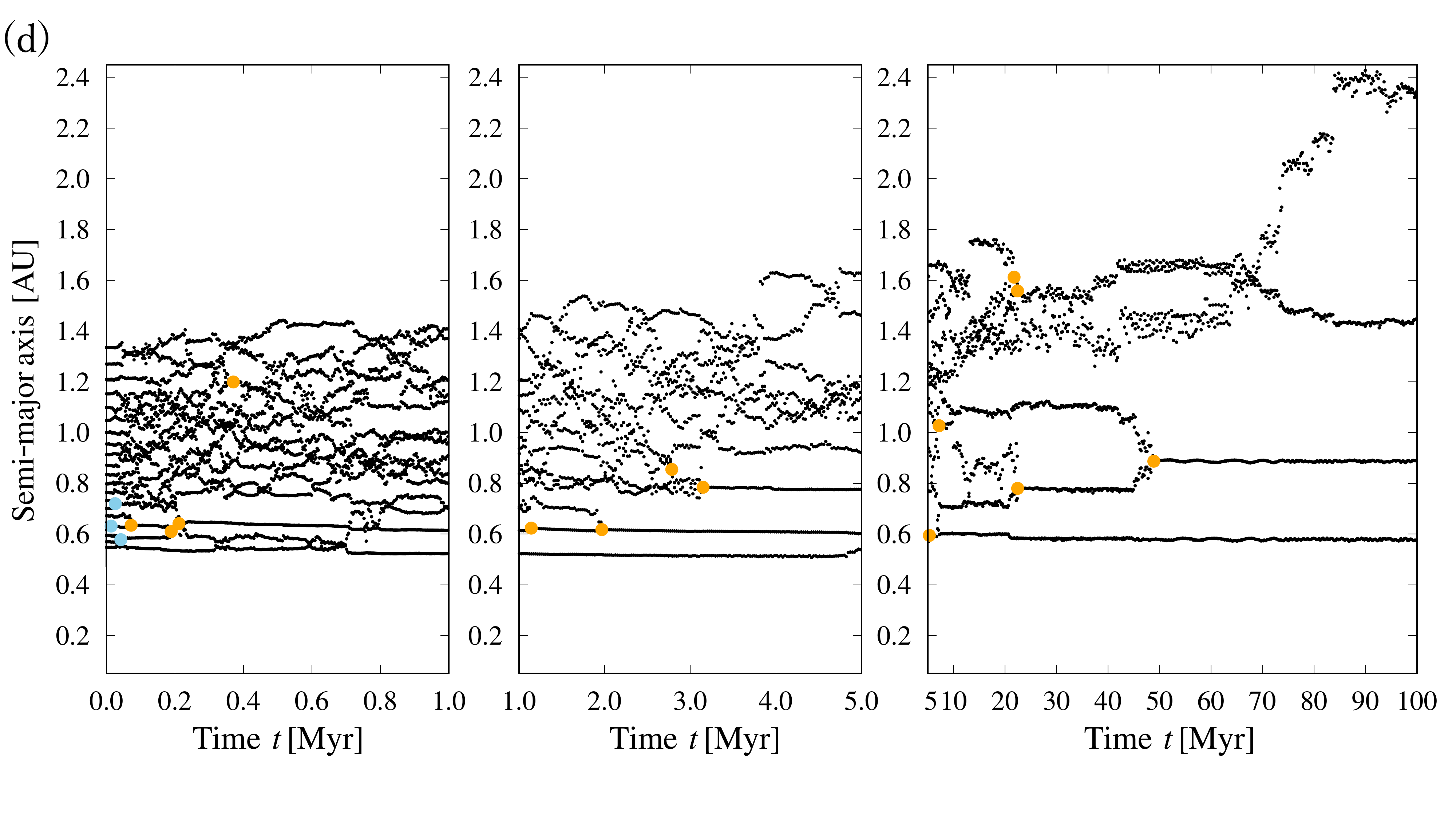}
\end{figure}

Figure~\ref{fig:orbital evolution for i not 0} shows the dynamical evolution of protoplanetary systems for different initial configurations. Each system initially contained 21 protoplanets. As the disk gas dissipates, the systems become dynamically unstable and giant impacts occur. In all cases, initial giant impacts occur within first $10^{6}$ years. Within the first million years, the number of protoplanets decreases by about half. However, giant impacts continue intermittently thereafter, and late giant impacts (after $\sim10^{7}$ years) can still form Earth-mass planets.

\begin{figure}[ht!]
\centering
\includegraphics[width=0.9\textwidth]{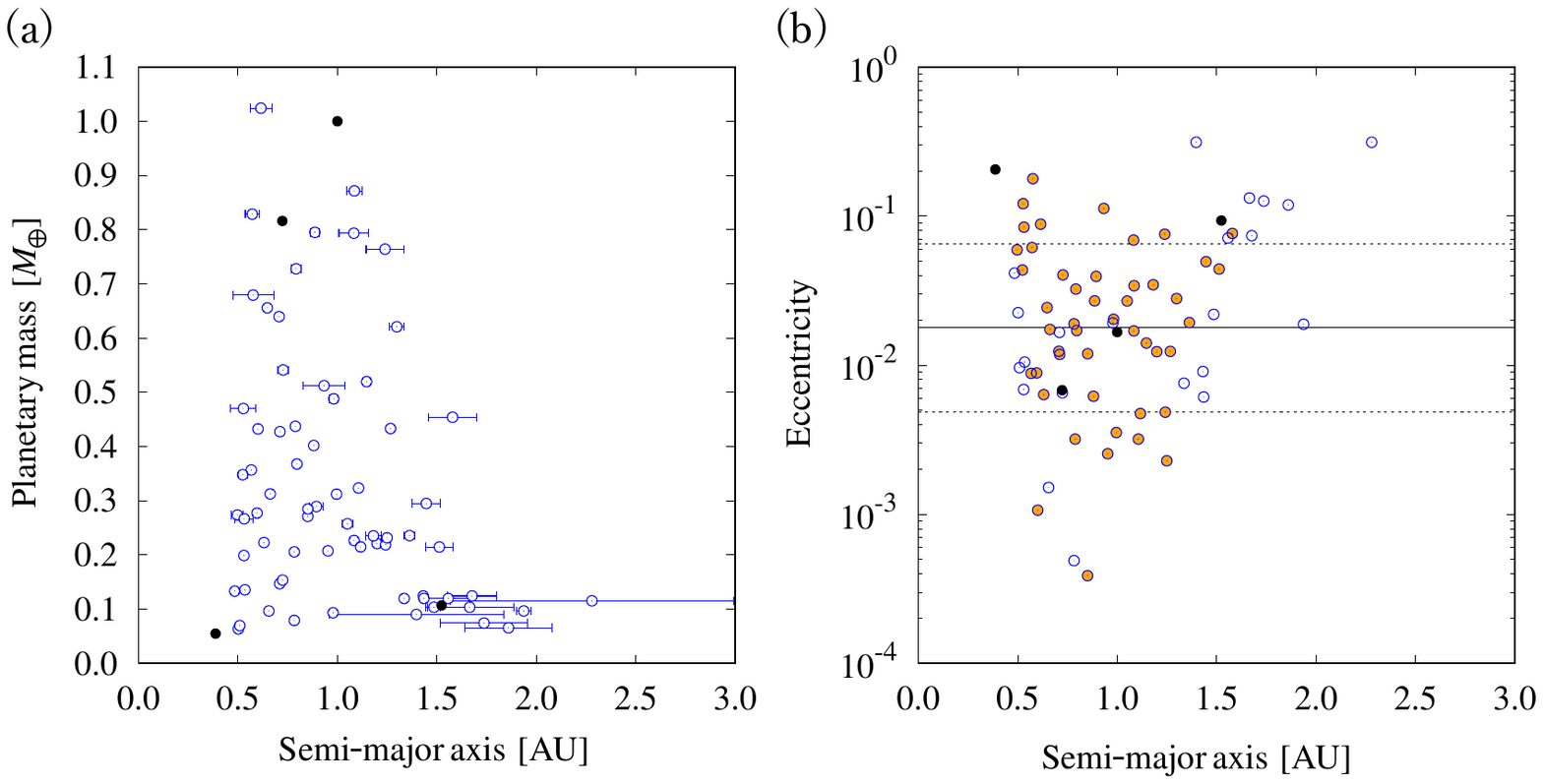}
\caption{A distribution of (a) planetary masses and (b) eccentricities for remaining planets at $t_{\mathrm{end}}$ in all calculations. The open circles are the result of our calculations, and the filled black ones are current rocky planets in the solar system. The filled orange circles for eccentricity plots represent planets with a mass of $0.2 \, M_{\oplus}$ or more. Error bars represent perihelion and aphelion. The solid line and dashed lines represent mean and standard deviation values of the common logarithm of eccentricity for planets with a mass of $0.2 \, M_{\oplus}$ or more. The distributions of the finally-formed planets roughly cover the current semi-major axis, mass, and eccentricity distributions of the rocky planets in the solar system. }
\label{fig:distribution of finally-formed planets}
\end{figure}

We show the distribution of orbital elements for the finally-formed planets. Figure \ref{fig:distribution of finally-formed planets} shows the distribution of planetary masses and eccentricities. The distributions of the finally-formed planets broadly cover the current semi-major axis, mass, and eccentricity distributions of the rocky planets in the Solar System. In our simulations, the minimum and maximum semi-major axes are $a_{\mathrm{min}} = 0.484$ AU and $a_{\mathrm{max}} = 2.28$ AU. Compared with the initial configuration, the final distribution of semi-major axes is more dispersed due to gravitational scattering (see Figure \ref{fig:initial_orbit} and Methods). The initial protoplanets were assigned masses of about 0.1 Earth masses, and the initial system was truncated at the endpoints for modeling convenience. For this reason, it is most appropriate to compare the distributions of the finally formed planets in this study with those of Earth and Venus. 
In terms of mass, planets comparable to Venus and Earth formed, although only 3 out of 69 planets exceeded the mass of Venus. For eccentricity, the mean and standard deviation of $\log_{10} e$ for planets with masses greater than $0.2 \, M_{\oplus}$ are $\langle \log_{10} e \rangle = -1.75$ and $\sigma_{\log_{10} e} = 0.56$. Earth and Venus fall well within the range $\langle \log_{10} e \rangle \pm \sigma_{\log_{10} e}$. In our results, more massive planets tend to have slightly higher eccentricities.

The distribution of eccentricities allows comparison of our results with not only rocky planets in the solar system, but also previous studies incorporating damping gas drag force. \citet{kominami02} did not perform a statistical analysis of the eccentricity, but under disk conditions similar to those of our study(the case of the dissipating disk), the planets analogous to Earth and Venus are formed respect to planetary mass, semi-major axis, and eccentricity. We also confirmed the Earth-sized planets (larger than Venus) with low eccentricity ($e < 0.1$) as well as \citet{kominami02} did. For example, finally-formed planets larger than Venus in our study have eccentricities of 0.034 - 0.088. On the other hand, \citet{kominami02} confirmed a system in which multiple planets with eccentricities of about 0.001 - 0.01 are formed in the rocky planet region under the disk condition similar to those in our study. These eccentricities are about one order of magnitude smaller than those of our results.

When examining the atmospheric accretion of protoplanets, particular attention must be paid to initial giant impacts, because by the time of late giant impacts, the disk gas has substantially dissipated compared to earlier time. In a disk with $\tau_{\mathrm{diss}} = 10^{6}$ yr, the gas surface density declines by about a factor of 150 within 5 Myr. Therefore, we first consider initial giant impacts to estimate atmospheric accretion, and then incorporate the effects of late giant impacts to capture the full evolutionary picture.

We now estimate the minimum level of disk dissipation required for a protoplanet to acquire sufficient hydrogen to reproduce Earth's present composition. Let $M_{\mathrm{{H}_{2}, required}}$ represent the hydrogen mass needed to match Earth's current atmospheric conditions. \citet{young23} showed that a $0.5\,M_{\oplus}$ protoplanet with 0.2 wt\% hydrogen in its atmosphere can yield Earth-like conditions:
\begin{equation}
M_{\mathrm{{H}_{2}, required}} = 0.5 \, M_{\oplus} \times 0.2\ \mathrm{wt}\% \approx 5.972 \times 10^{24}\, \mathrm{g}.
\label{eq:gas required}
\end{equation}
If $M_{\mathrm{atm}}$ is the mass of atmosphere acquired from the disk when it has dissipated to a fraction $\kappa$ of the MMSN value, then for a protoplanet at 1 AU, $\rho_{\mathrm{gas}} = \kappa \, \rho_{\mathrm{gas}}^{\mathrm{min}}$ and $\Sigma_{\mathrm{gas}} = \kappa \, \Sigma_{\mathrm{gas}}^{\mathrm{min}}$.

Setting $M_{\mathrm{{H}_{2}, required}} = M_{\mathrm{atm}}$ yields $\kappa$ value and a time when a giant impact should occur $t_{\mathrm{col}}$ depending on the three models of atmospheric mass $M_{\mathrm{torus}} < M_{\mathrm{mid}} < M_{\mathrm{cylinder}}$. The estimates are summarized in Table \ref{tab:estimate of kappa}.

\begin{table}[h]
\begin{center}
\caption{Atmospheric mass models and gas dissipation}
\begin{tabular*}{110mm}{@{\extracolsep{\fill}}ccc} \hline
   $M_{\mathrm{atm}}$ & $\kappa$ & $t_{\mathrm{col}}$ \\
   \hline
    $M_{\mathrm{torus}}$ & 1/697 & $1.942 \times 10^{6}$ yr \\
    $M_{\mathrm{mid}}$ & 1/3182 & $3.460 \times 10^{6}$ yr \\
    $M_{\mathrm{cylinder}}$ & 1/6356 & $4.151 \times 10^{6}$ yr \\
\hline
\end{tabular*}
\end{center}
\tablecomments{Gas dissipation rates and times when a giant impact should occur to reproduce the results of \citet{young23}. The merged protoplanet has a mass of $0.5\,M_{\oplus}$ and is located at 1 AU.}
\label{tab:estimate of kappa}
\end{table}

Thus, ideally, giant impacts should occur when the disk gas has dissipated to about 1/6400 - 1/700 of the MMSN value for a $0.5\,M_{\oplus}$ protoplanet to acquire 0.2 wt\% hydrogen.

Under the disk conditions assumed in this study ($\tau_{\mathrm{diss}} = 10^{6}$ yr and $\Sigma_{\mathrm{gas}}(0) = 0.01\,\Sigma_{\mathrm{gas}}^{\mathrm{min}}$), $\kappa$ corresponds to a collision time of $t_{\mathrm{col}} \simeq 1.9 \,\,\text{-}\,\, 4.2 \times 10^{6}$ yr for a merged protoplanet with $M_{\mathrm{proto}} = 0.5\,M_{\oplus}$. However, as shown in Figure~\ref{fig:orbital evolution for i not 0}, giant impacts occur across a wide range of timescales, from early collisions in a dense gas disk to late impacts in a nearly dissipated disk. In particular, most giant impacts occur before the gas disk has significantly dissipated, making it challenging to achieve the precise timing required to replicate Earth's composition. In this context, it is important to trace the entire formation process, including both initial giant impacts in a dense gas disk and later impacts in a nearly dissipated disk.

\begin{figure}[ht!]
\centering
\includegraphics[width=0.9\textwidth]{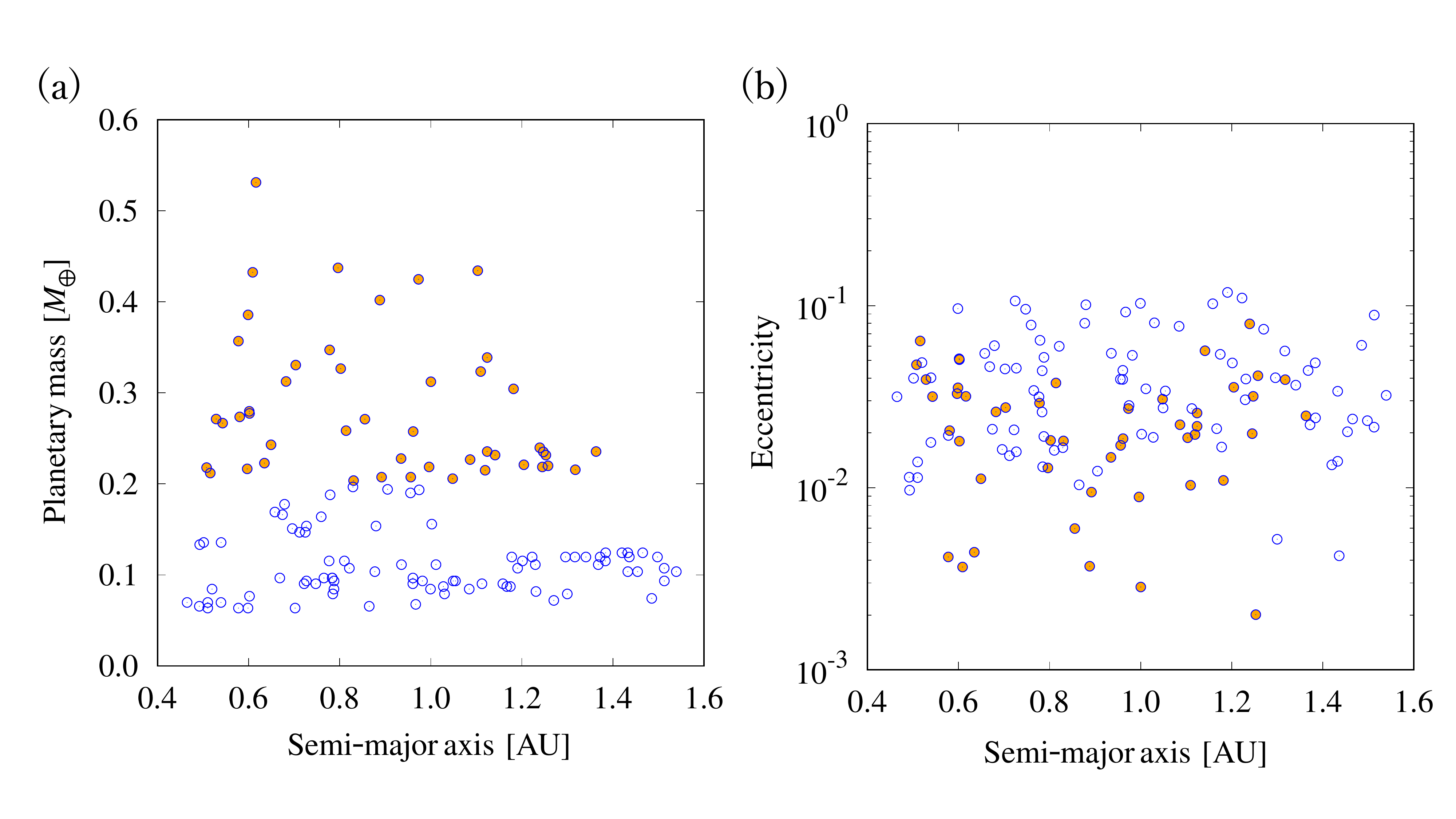}
\caption{A distribution of (a) planetary masses and (b) eccentricities for remaining protoplanets at $t = 1.6 \times 10^6$ yr in all calculations. The filled circles in orange represent protoplanets with a mass of $0.2 \, M_{\oplus}$ or more. These protoplanets are assumed to acquire hydrogen-rich atmosphere and incorporate it in thir interior. Hydrogen-rich protoplanets are distributed evenly relative to the total distribution.}
\label{fig:distribution of protoplanet at 1e7 GPLUMyr}
\end{figure}

Finally, we show how many initial giant impacts that supply protoplanets with excess amounts of hydrogen and its distributions. Figure \ref{fig:distribution of protoplanet at 1e7 GPLUMyr} shows distributions of protoplanets with a mass of $0.2 \, M_{\oplus}$ at $t = 1.6 \times 10^{6}$ yr. At this time, for all three models of atmospheric mass, the protoplanet is expected to take in more hydrogen by giant impact than the amount that reproduces the present-day Earth. At the time, the average number of protoplanets in the system is 11.0, and of these, the average number of protoplanets with $0.2 \, M_{\oplus}$ or more is about 3.92 (detail in Table \ref{tab:inclination N-body}).

\subsection{Results of geochemical equilibration} \label{sec:result_geochem}

Further analysis of chemical equilibrium calculations is crucial for determining whether protoplanets can ultimately reproduce Earth's chemical composition. If the initial giant impacts occurs while a substantial amount of disk gas remains, protoplanets would acquire an excessive hydrogen-rich atmosphere, preventing them from replicating Earth's composition. Conversely, if the gas density falls much below 1/6400 - 1/700 of the MMSN, protoplanets cannot acquire sufficient hydrogen to reproduce Earth's conditions. In most cases, the system experiences initial giant impacts significantly earlier than the time when the gas density matches the value needed to reproduce Earth. This suggests that initial giant impacts alone could not create protoplanets with an Earth-like surface environment (as discussed in more detail in Section~\ref{sec:result_geochem}).

Before incorporating $N$-body simulation results into chemical equilibrium calculations, we conducted a preliminary investigation of the chemical equilibrium behavior over a broader range of hydrogen atmospheres than considered in previous studies such as \citet{young23}. It was previously shown that a protoplanet with a 0.2 wt\% hydrogen atmosphere and a mass of 0.5\,$M_{\oplus}$ can reproduce Earth's chemical composition through a series of chemical reactions.

\begin{figure}[ht!]
\plotone{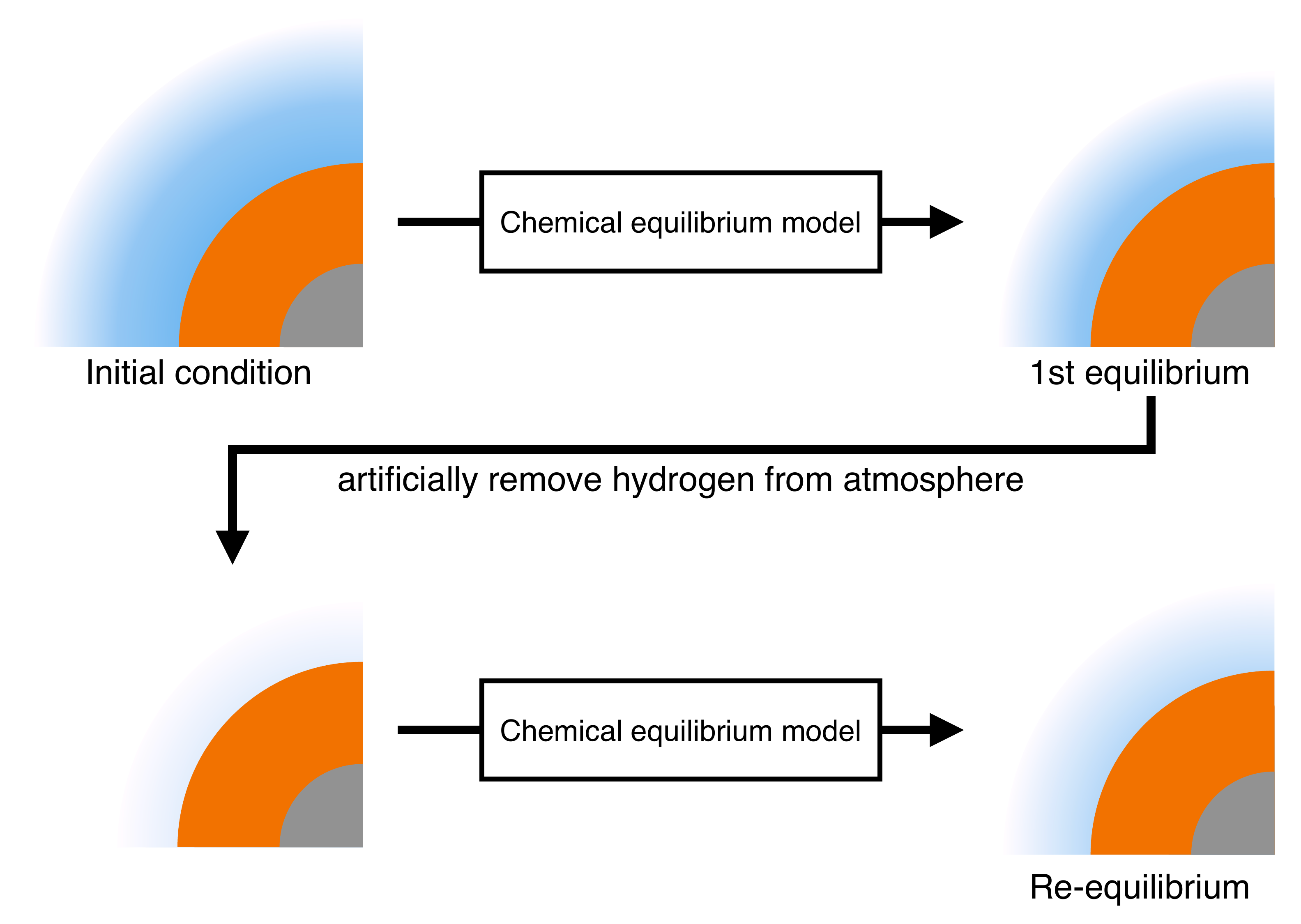}
\caption{Chemical re-equilibration process. After the initial equilibrium calculation, all hydrogen in the atmosphere is artificially removed, and equilibrium is recalculated.}
\label{fig:re_equilibrium_process}
\end{figure}

As demonstrated by \citet{genda05}, a protoplanet can lose part of its previously retained atmosphere during a giant impact event, especially if a surface liquid ocean is present. Therefore, to model an alternative scenario, we conducted chemical equilibrium calculations assuming that, after initially reaching equilibrium under a hydrogen atmosphere, the atmosphere undergoes selective removal of hydrogen during a giant impact, followed by re-equilibration of the system (Figure~\ref{fig:re_equilibrium_process}).

\begin{figure}[ht!]
\plotone{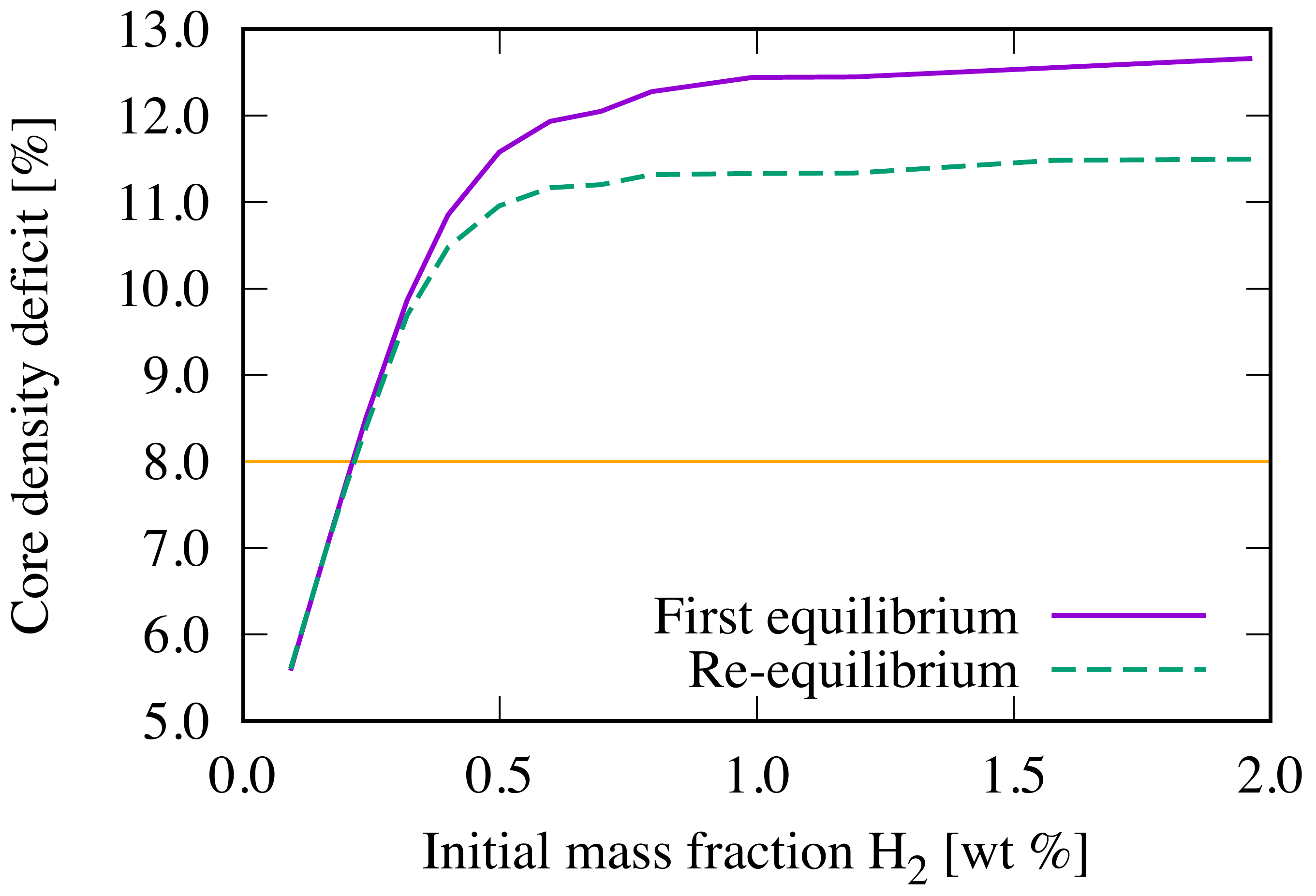}
\caption{Relationship between the initial amount of hydrogen atmosphere and the core density deficit. The dashed line shows the results after re-equilibration: hydrogen is artificially removed from the atmosphere after the initial equilibrium calculation, and equilibrium is recalculated. An 8\% density deficit (orange line) corresponds to the present-day Earth. As the initial hydrogen content increases, more hydrogen enters the core, but this effect saturates. Re-equilibration returns some hydrogen from the core back to the atmosphere. Our first equilibrium calculations follow \citet{young23}, although they explored only a narrow range of initial hydrogen mass fractions (near 0.2 wt\%).}
\label{fig:H2-core_DD}
\end{figure}

Our analysis revealed that the chemical equilibrium state of a protoplanet is sensitive to the amount of hydrogen in the atmosphere, up to approximately 0.5 wt\% (Figure~\ref{fig:H2-core_DD}). Beyond this threshold, the chemical response becomes insensitive, suggesting a saturation tendency. Although the exact cause of this saturation is unclear, it is unlikely that hydrogen can continue penetrating into the iron core without limitation.

The outcome of the re-equilibration process varied depending on the initial amount of hydrogen atmosphere (see Figure~\ref{fig:H2-core_DD}). When the initial hydrogen content was low, a significant fraction of hydrogen entered the iron core. If the initial atmosphere contained less than about 0.3 wt\% hydrogen, removing hydrogen and re-equilibrating had little effect on the chemical composition, because most hydrogen had already migrated into the core. In contrast, when the initial hydrogen atmosphere was high, a considerable amount of hydrogen remained in the atmosphere. Subsequent giant impacts removed this excess hydrogen, shifting the chemical equilibrium and causing some hydrogen originally incorporated into the iron core to return to the atmosphere. When the initial hydrogen content exceeded approximately 0.3 wt\%, it was confirmed that hydrogen once incorporated into the iron core could be partially returned to the atmosphere through re-equilibration. Although the amount of hydrogen that can be transferred back is limited, re-equilibration effectively restored the core density deficit by approximately 1--2\%. Nevertheless, once the system is saturated, it is difficult to reduce the core density deficit to the current level of the Earth by this re-equilibrium. In the context of planet formation, this corresponds to a late giant impact after disk gas dissipation between protoplanets that are both saturated by prior initial impacts.

Overall, these results imply that hydrogen inventories in protoplanetary cores can vary substantially depending on the timing and frequency of re-equilibration events. By combining $N$-body simulations with chemical equilibrium models, we capture a more dynamic and realistic picture of how late impacts contribute to the final water budget and core density deficit of terrestrial planets.

We discuss the robustness of the chemical calculation for planetary mass and surface pressure in Section \ref{discussion:robustness of chemical calculation}.

\subsection{Integration of N-body calculation and geochemical equilibration} \label{sec:integration}

In this section, we integrate the results of the $N$-body simulations into the chemical equilibrium calculations. The amount of hydrogen supplied to each protoplanet was determined based on the giant impact timings obtained from the $N$-body simulations, and a series of chemical equilibrium calculations was performed accordingly. The initial chemical equilibrium calculation following the first giant impact remained consistent with that of \citet{young23}, with adjustments made only to the atmospheric hydrogen content. For subsequent giant impacts, the initial conditions for the chemical equilibrium calculations were derived from the outcomes of the previous equilibrium states. As described in Section~\ref{sec:result_geochem}, the hydrogen content was updated based on the timing of each giant impact. If the disk gas had already dissipated by the time of a later impact and a new hydrogen atmosphere could not be acquired, hydrogen previously captured in the iron core during earlier impacts could be released back into the atmosphere. This reflects the reversible nature of hydrogen transfer between the atmosphere and the iron core.

As discussed above, giant impacts among protoplanets occur intermittently, with initial and late giant impacts playing distinct roles (see Figure~\ref{fig:orbital evolution for i not 0}). During the initial giant impacts, a substantial amount of disk gas remains around the protoplanets. Consequently, after each giant impact, the protoplanets can acquire a large amount of atmosphere, leading to significant hydrogen absorption into their cores. This results in core density deficits that are much larger than that of present-day Earth. After the initial giant impacts subside, the frequency of giant impacts decreases substantially (over timescales of at least $\sim 10^{7}$ years). During this interval, most of the disk gas dissipates (for $\tau_{\mathrm{diss}} = 10^{6}$ years, the gas surface density decreases by a factor of $e^{-10}$ over $10^{7}$ years).

During the late giant impacts, protoplanets may lose their preexisting atmospheres but are unable to reacquire significant amounts of gas from the disk because little material remains. In these cases, chemical reactions within the protoplanets fall into three main categories:  
(1) a giant impact between two protoplanets, both possessing hydrogen-rich cores;  
(2) a collision between a protoplanet with a hydrogen-rich core and one with a pure-iron core;  
(3) a collision between two protoplanets with pure-iron cores.  
Specifically, in case (1), hydrogen previously dissolved in both cores can partially re-equilibrate and return to the atmosphere; in case (2), the hydrogen concentration is diluted through merging, reducing the overall core density deficit; and in case (3), no significant change occurs in chemical composition because neither core contains hydrogen.

The extent to which hydrogen incorporated into the iron core during the first chemical equilibrium returns to the atmosphere during the second chemical equilibrium depends on the initial conditions of the first equilibrium. If the protoplanet acquires a moderate atmosphere during the first equilibrium—one that reproduces present-day Earth conditions—then the hydrogen incorporated into the core will not return to the atmosphere during the second equilibrium. This is because the amount of hydrogen remaining in the atmosphere after the first equilibrium calculation is minimal. Even if this residual hydrogen is removed by a subsequent giant impact, its effect on the second equilibrium is negligible. In contrast, when excess hydrogen is incorporated into the iron core during the first chemical equilibrium and saturation occurs, some hydrogen can return to the atmosphere during the second equilibrium. All protoplanets that experienced giant impacts while the disk gas was still dense developed hydrogen-rich cores relative to modern Earth.

\begin{figure}[ht!]
\plotone{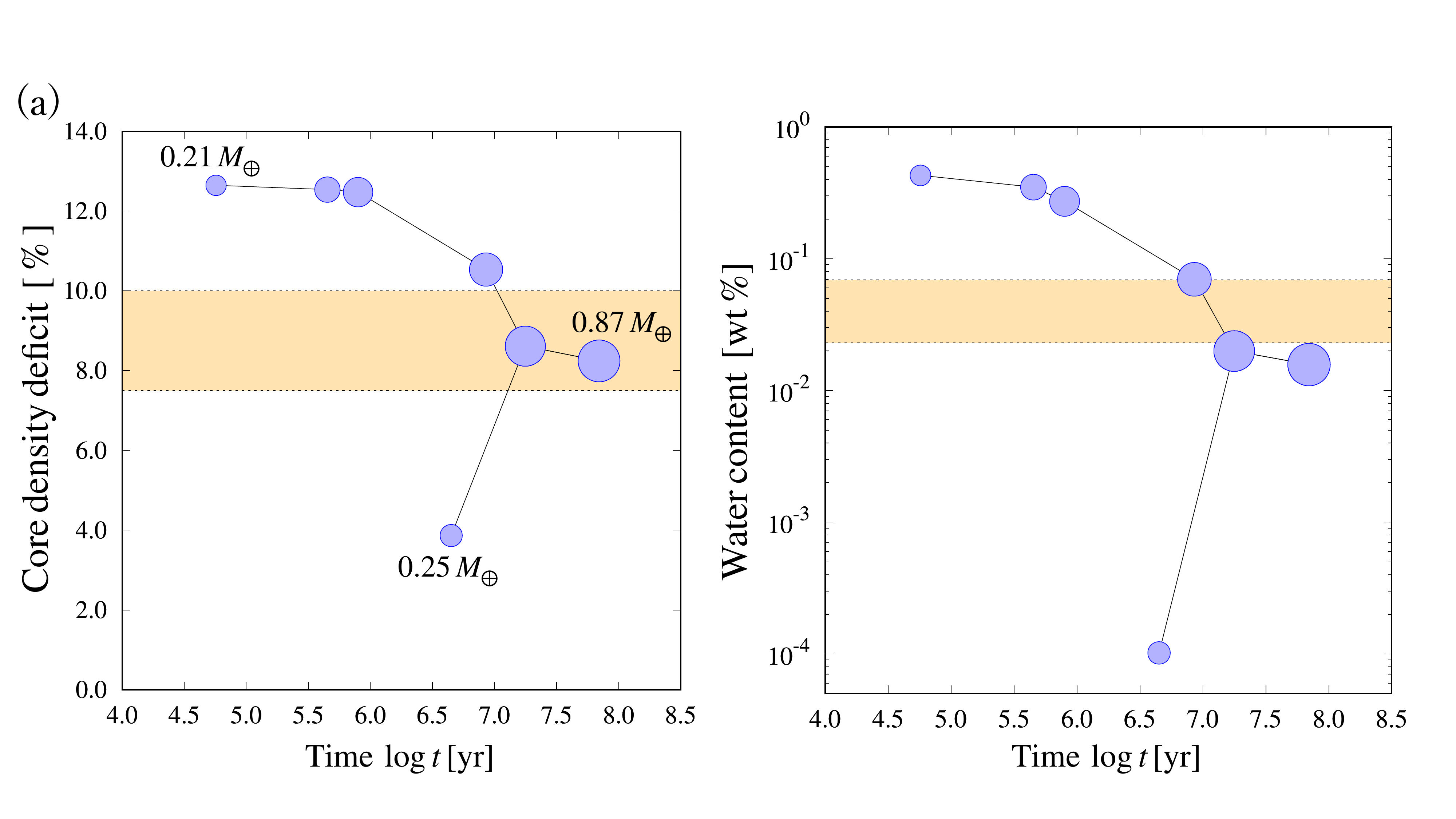}
\caption{Evolution of chemical compositions of some planets. This results correspond a atmospheric mass of $M_{\mathrm{torus}}$. We show the evolution of (a) third planet in Case No.6 and (b) second planet in Case No.12 at $t_{\mathrm{end}}$. The horizontal axis represents time, and each point corresponds to a protoplanet immediately after a giant impact, with the symbol area proportional to its mass. 7.5 - 10.0 \% of core density deficit and 0.023 wt \% of water content represent the current Earth's value. For water content, we assume that 1 - 3 ocean mass can reproduce the current surface ocean. Protoplanets incorporate a large amount of hydrogen into their cores during initial giant impacts ($t \lesssim 10^{6} \, \mathrm{yr}$). These protoplanets have too large core density deficit to reproduce the current Earth. However, through late giant impacts (after gas dissipation), collisions between hydrogen-rich and hydrogen-poor protoplanets can yield final planets with moderate core density deficits similar to Earth.
\label{fig:form of chemical composition}}
\end{figure}

\begin{figure}[ht!]
\plotone{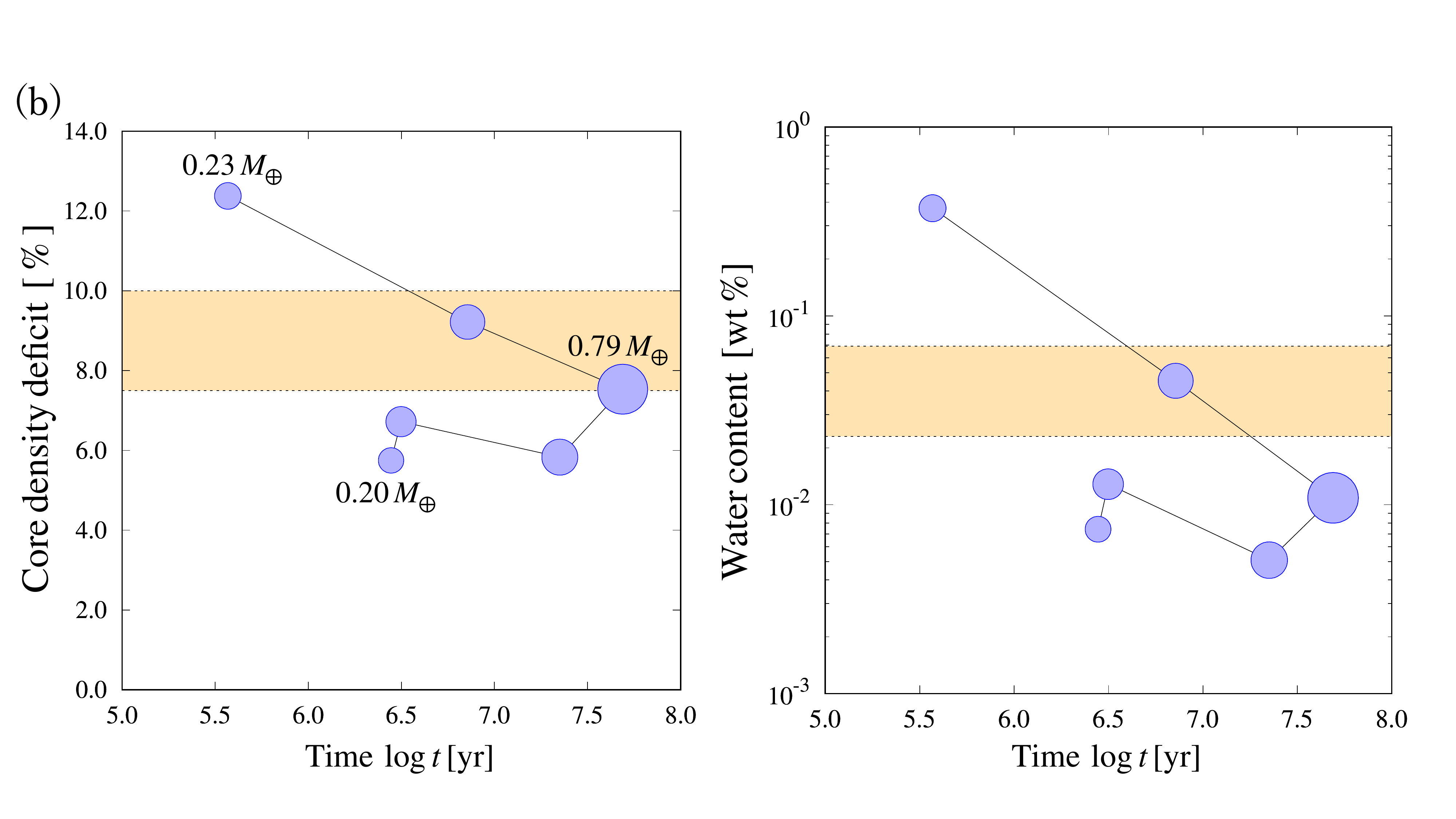}
\end{figure}

Even if substantial hydrogen is incorporated into the core during the initial giant impacts, the return of hydrogen to the atmosphere has only a limited effect in reducing excessive core density deficits. Re-equilibration can decrease the deficit by at most about 1\% (see Figure~\ref{fig:H2-core_DD}). Thus, to produce a final planet with a core density deficit comparable to that of Earth, it is necessary for a hydrogen-rich protoplanet to collide with one that retains a nearly pure-iron core. In one of our simulations, such a collision successfully formed two Earth-like protoplanets with core density deficits matching the modern Earth (Figure~\ref{fig:form of chemical composition}).

In these calculations, we assumed that only light hydrogen is lost during large impacts, while heavier water vapor remains in the atmosphere \citep{genda05}. Consequently, the water vapor content of the protoplanet does not change significantly even after multiple giant impacts. When hydrogen is released from the core back into the atmosphere, it is assumed to contribute to water formation.

For a criteria of core density deficit and water content value, we set ranges that could be consistent with the current Earth, based on previous studies. For core density deficit, we use some results of high temperature and high pressure experiments (See Section \ref{Section:Formation of Earth-like planet}) and set 7.5 - 10.0 \% as a range to reproduce the current Earth. For water content, \citet{KUROKAWA2018149} modeled a global cycling and evolution of water on Earth taking the D/H compositions into account and calculated the difference in D/H ratio between the surface and interior to give the constraint that the volume of initial oceans could be 2 to 3 times larger than that of current Earth. Here, we set 1 to 3 ocean mass as a range of water content. Regarding water content, we used the surface ocean mass of present-day Earth or its formation time as the reference value. However, it is known that Earth's interior also contains a substantial amount of water. High-pressure experiments have shown that transition zone minerals can store up to about 3 wt\% water within their crystal structures \citep{Inoue1995}. This suggests that the amount of water stored in Earth's mantle is much greater than that on the surface. Therefore, surface and atmospheric water behavior during giant impacts may need to be investigated in more detail.

\begin{figure}[ht!]
\plotone{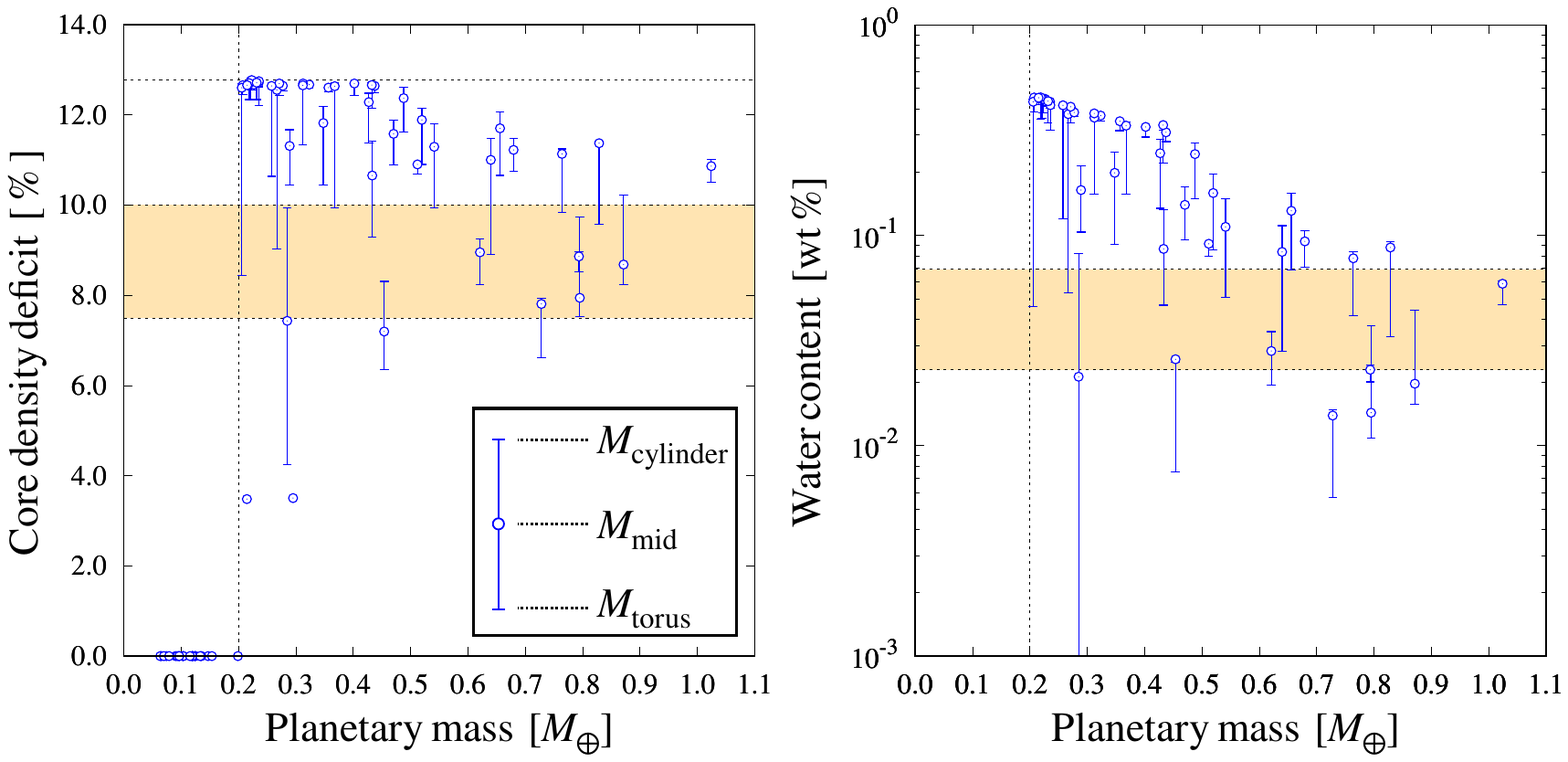}
\caption{A distribution of core density deficits and atmospheric water contents for remaining planets at $t_{\mathrm{end}}$ in all
calculations. Open circles correspond to results with an atmospheric mass of $M_{\mathrm{mid}}$, and error bars correspond that of $M_{\mathrm{torus}}$ and $M_{\mathrm{cylinder}}$. Planets with masses less than $0.2 M_{\oplus}$ have no core density deficit, since we conducted chemical equilibrium calculation for only that greater than $0.2 M_{\oplus}$ (Method). Core density deficit of 7.5-10.0 \% represents a range consistent with the current Earth and about 12.7 \% is saturated value. Low-mass planets formed in early giant impacts acquire a large amount of disk gas and get a high core density deficit. As the planetary mass increases due to subsequent giant impacts, the core density deficit decreases and approaches a value consistent with that of the current Earth. As with core density deficit, water content also decreases as planetary mass increases.
\label{fig:distribution of core DD}}
\end{figure}

Finally, we show the distribution of core density deficit for finally-formed planets. Figure \ref{fig:distribution of core DD} shows the relationship between planetary mass and core density deficit. First, Error bars represent the effect of atmospheric mass on core density deficit for each planet. Larger atmospheric mass results in higher core density deficit. Second, core density deficit decreases slowly with planetary mass. The distribution of water contents behaves similarly to the distribution of the core density deficits. 

Points without error bars in the core density deficit panel correspond to protoplanets that experienced their first giant impacts only after disk dissipation. In such cases, the protoplanets accreted no atmosphere, so $M_{\mathrm{atm}} \simeq 0$ in all three atmospheric mass models, although the cores still incorporated Si and O from the mantles, leading to slight density deficits. The atmospheric mass $M_{\mathrm{mid}}$ is defined as the average of $M_{\mathrm{torus}}$ and $M_{\mathrm{cylinder}}$, but many $M_{\mathrm{mid}}$ points do not lie midway between the corresponding $M_{\mathrm{torus}}$ and $M_{\mathrm{cylinder}}$ values. This arises because the response of the calculated chemical compositions to atmospheric mass is non-uniform (see Figure \ref{fig:H2-core_DD}). For example, if $M_{\mathrm{mid}}$ lies at the transition between a linearly increasing regime and a saturated regime, the resulting density deficit is closer to that for $M_{\mathrm{cylinder}}$ than to that for $M_{\mathrm{torus}}$.

\section{Discussions} \label{sec:discussion}

\subsection{Formation of Earth-like planet}
\label{Section:Formation of Earth-like planet}

When comparing Earth with the planets formed in our simulations, we primarily consider the reproducibility of the core density deficit as the main criterion. We do not emphasize water production here, because Earth's water was likely supplied from multiple sources—including water-rich comets—reproducing Earth's water content solely through nebular gas capture is not necessary. Earth's core is approximately 10\% less dense than pure iron would be under similar conditions \citep{birch1964}, corresponding to about an 8\% density deficit in the profile of a $0.5\,M_{\oplus}$ protoplanet \citep{young23}. More recent high-pressure and high-temperature experiments suggest an outer-core density deficit of about 7.5 – 7.6\% \citep{Kuwayama2020}. These findings indicate that the target range for the core density deficit in our model should be around 7.5 - 10\%, consistent with Earth's inferred properties. We define an ``Earth-like planet'' as one with a mass similar to that of Earth and whose core density deficit falls within the observationally constrained ranges. We don't take the semi-major axis into account unless the planet is significantly outside the rocky planet region.

Our focus is primarily on the chemical composition of planets forming near 1 AU, rather than on their exact masses or orbital elements. This is because the initial 21 protoplanets in our model are larger than Mars-mass bodies. For instance, the outermost protoplanet has $M_{\mathrm{proto}} \sim 0.12\,M_{\oplus}$ at $\sim$1.3 AU ($e<0.01$), whereas Mars has a mass of $\sim0.1\,M_{\oplus}$. Consequently, we emphasize whether a planet forming near Earth's orbit can attain an ``Earth-like'' chemical composition. Future work aiming to distinguish Earth and Venus analogs more precisely will require additional orbital constraints.

Following \citet{young23} (and referencing \citealt{birch1964}), our results confirm that initial giant impacts can deliver excessive hydrogen into a protoplanet's core. Recent high-$P$-$T$ experiments simulating Earth's core–mantle boundary conditions have refined estimates of hydrogen abundance in the core. For example, \citet{Tagawa2021} inferred that $0.3$–$0.6\,\mathrm{wt\%}\,\mathrm{H}_2$ may reside in Earth's core, depending on the pressure–temperature path and element partitioning. It is important to note that \citet{young23} express hydrogen content relative to the protoplanet's total mass (0.2 wt\%), while \citet{Tagawa2021} define it relative to the core mass. In our model, protoplanets are approximately 65\% mantle and 35\% core by mass. Thus, if all the 0.2 wt\% hydrogen from \citet{young23} were incorporated into the core, the corresponding hydrogen concentration relative to the core would be $\sim0.57\,\mathrm{wt\%}$ ($0.2\,\mathrm{wt\%} \times (1/0.35)$), which falls well within the range inferred by \citet{Tagawa2021}. Combining mineral physics, geochemical, and cosmochemical data suggests that Earth's core likely also contains elements such as Si, O, and S \citep{Hirose2021}.

\begin{figure}[ht!]
\plotone{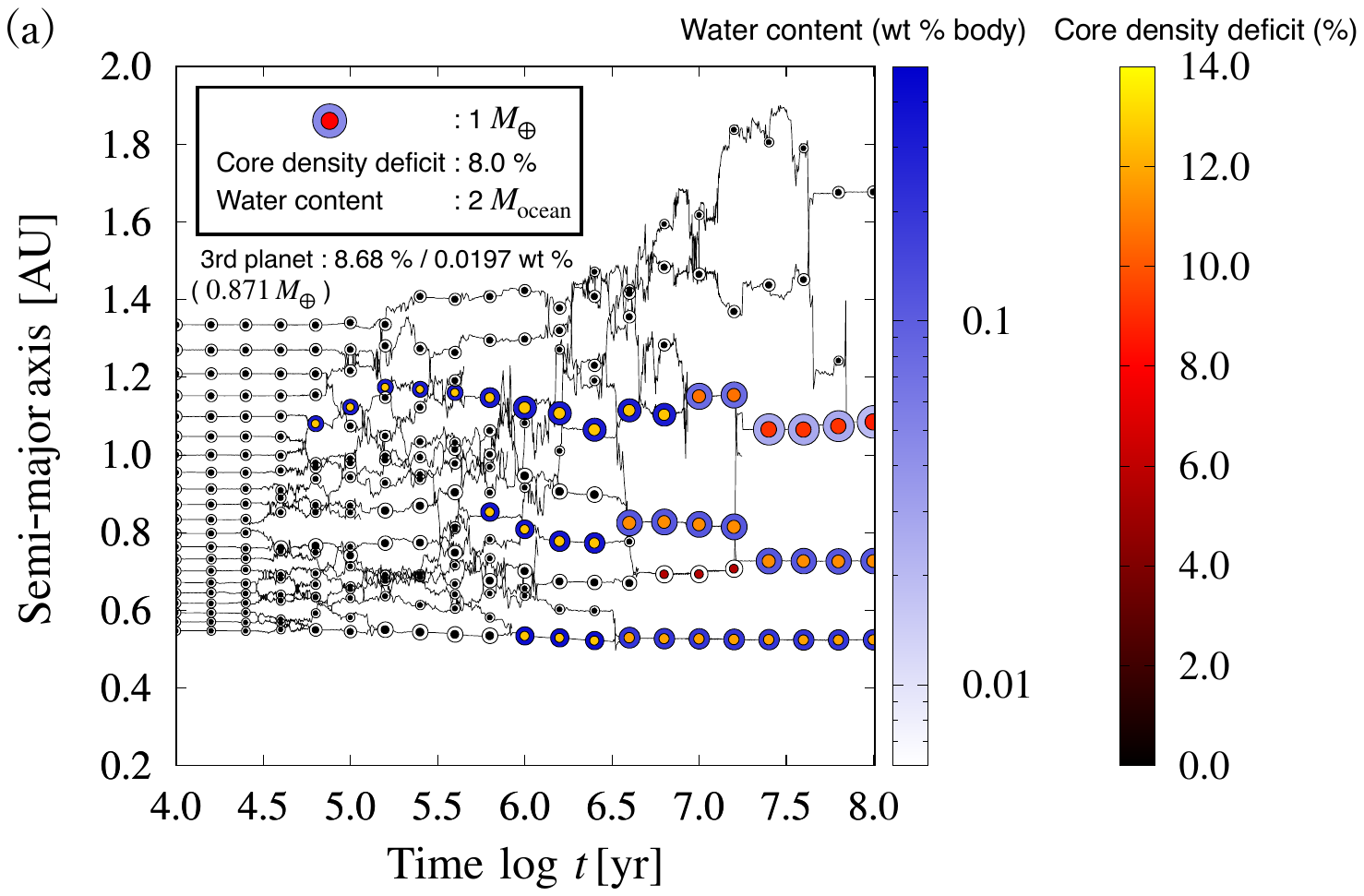}
\caption{Orbital evolution of protoplanets and changes in core density deficit and water content in atmosphere for (a) Case No.6, (b) No.7, (c) No.11, (d) No.12 (see Table~\ref{tab:inclination N-body}). We show results with an atmospheric mass of $M_{\mathrm{mid}}$. Early giant impacts in the inner region do not lead to atmospheric acquisition because the merged mass remains below the threshold for atmospheric capture ($M_{\mathrm{proto}}=0.2\,M_{\oplus}$). Even in the presence of inclination, the core density deficit evolves in a multistage manner. As cases where planets with a value close to the Earth's core density deficit was obtained, a planet with a final mass of $0.871\,M_{\oplus}$ and a core density deficit of 8.68\% forms at a final orbit of 1.085 AU(third planet in No.6), and a planet with a final mass of $0.795\,M_{\oplus}$ and a core density deficit of 7.95\% forms at a final orbit of 0.887 AU(second planet in No.12).
\label{fig:chemical_evolution_3D_No6}}
\end{figure}

\begin{figure}[ht!]
\plotone{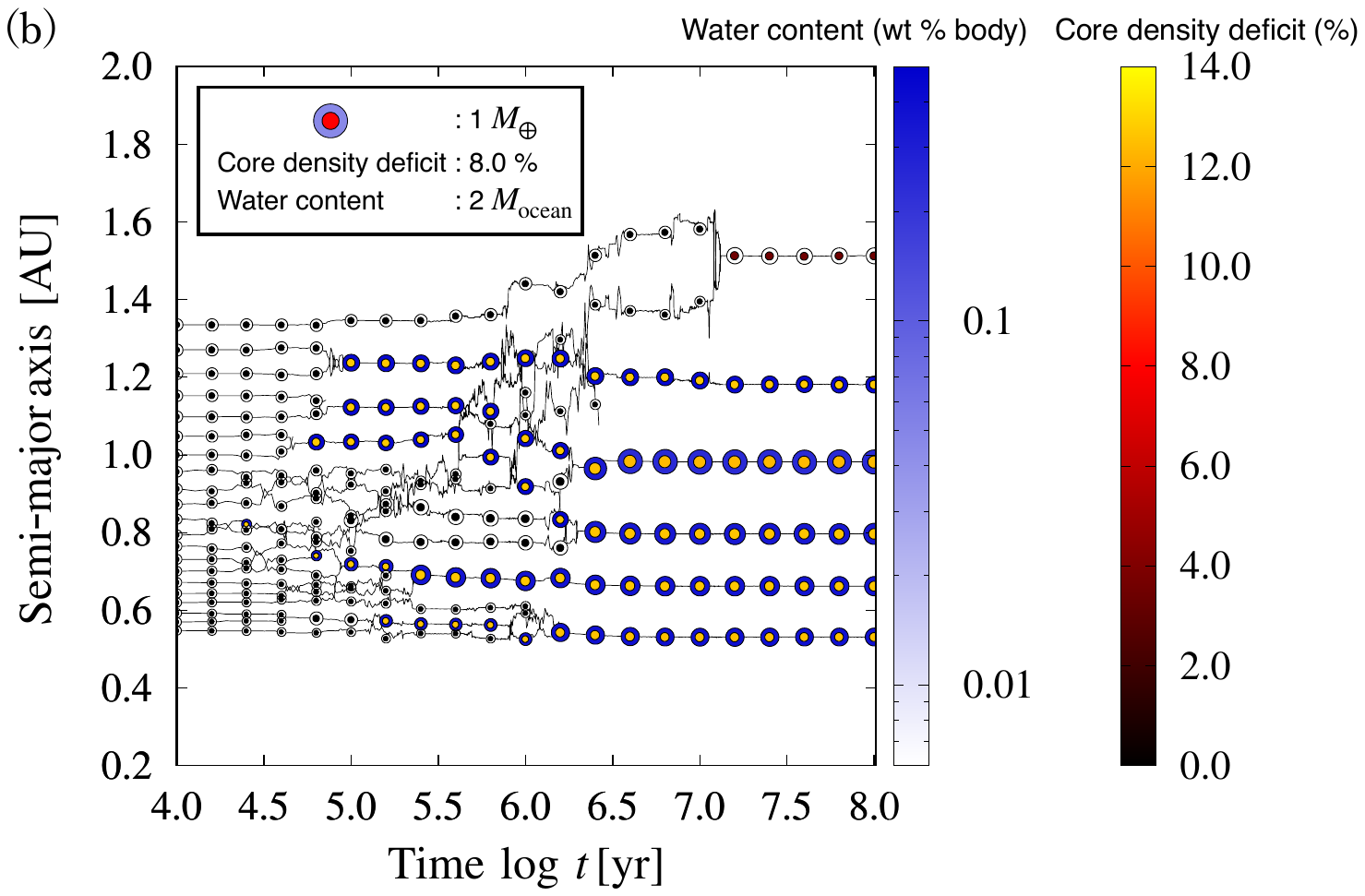}
\end{figure}

\begin{figure}[ht!]
\plotone{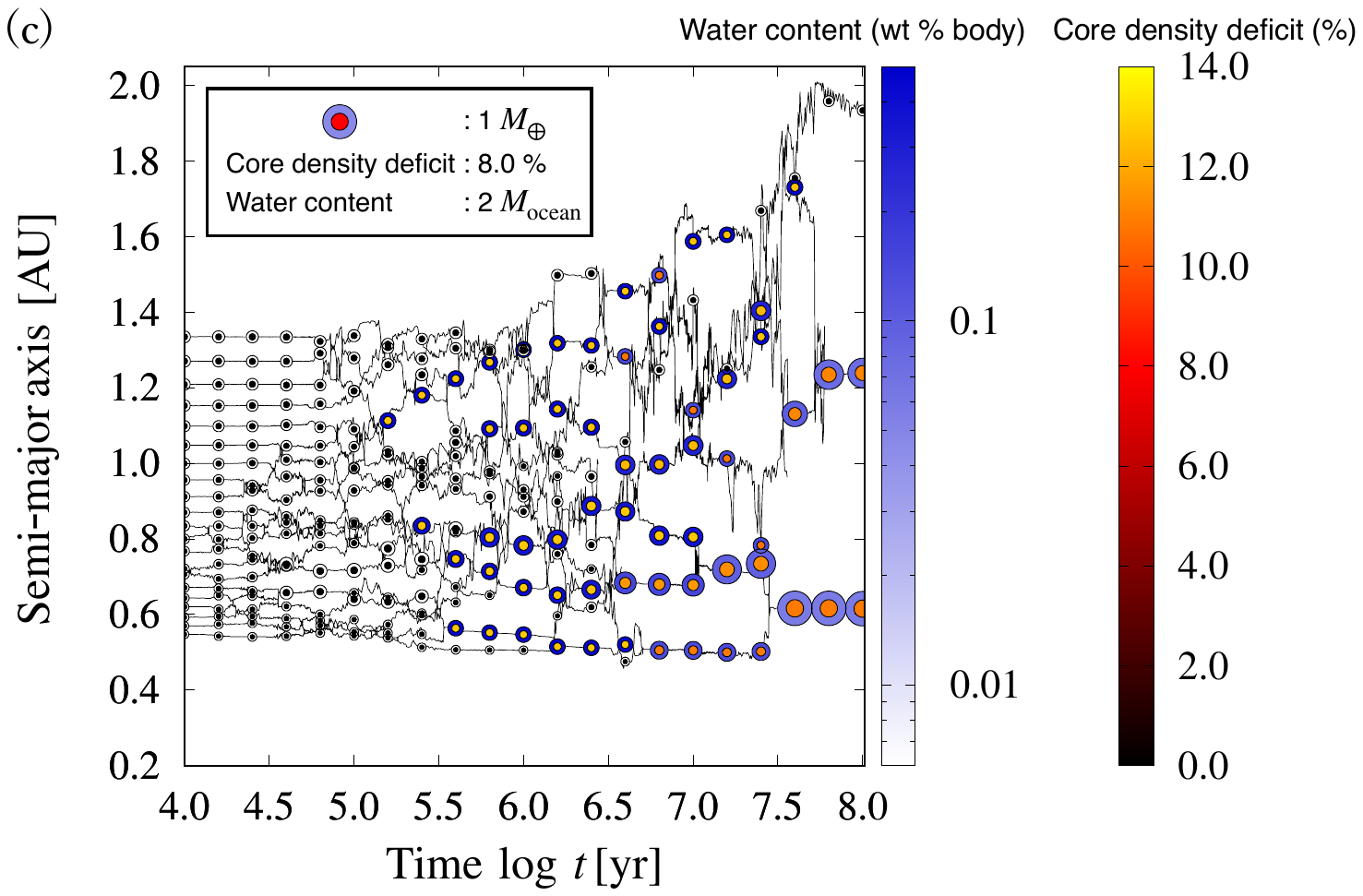}
\plotone{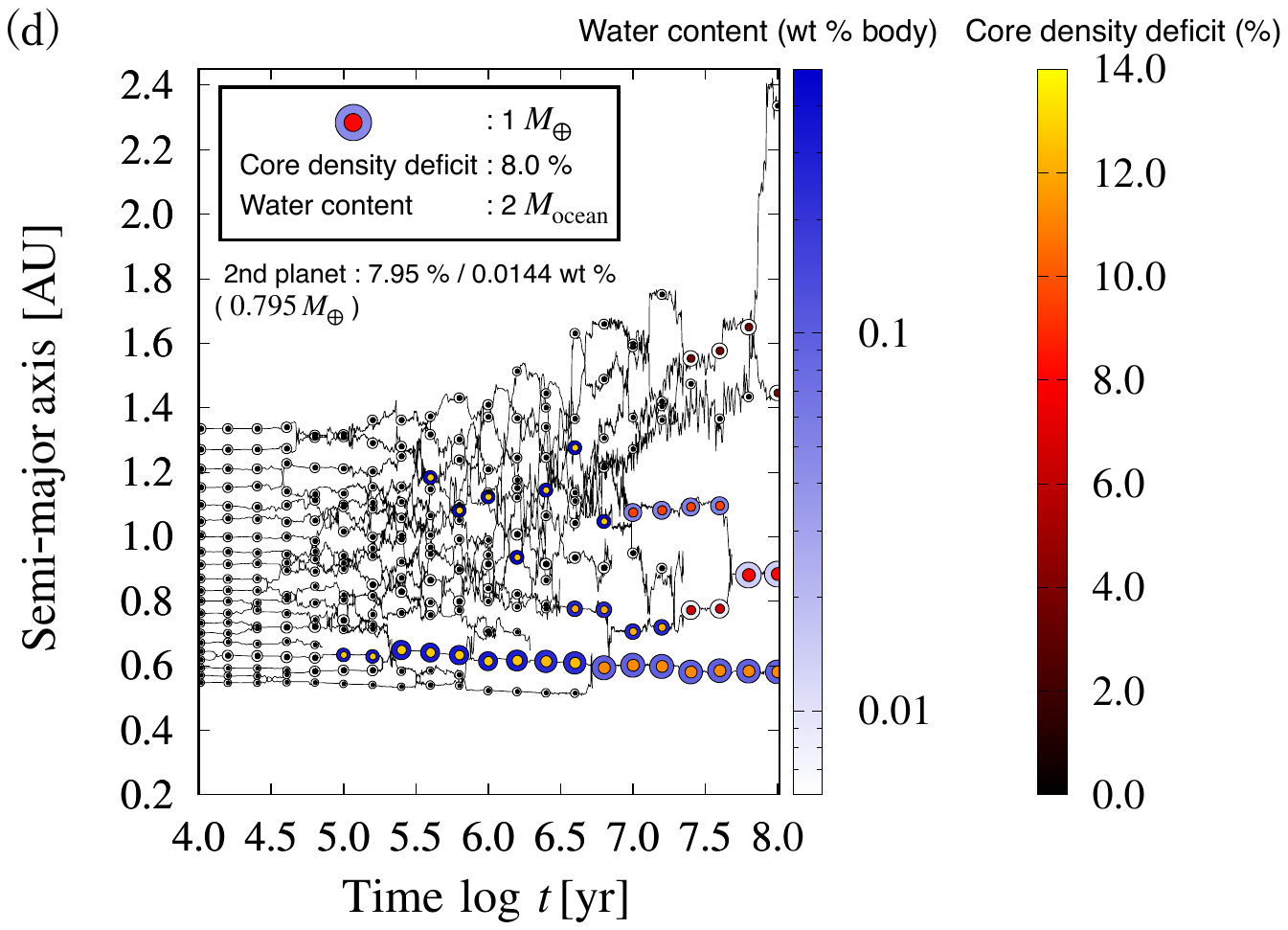}
\end{figure}

\begin{table}[h]
\begin{center}
\caption{Giant impact history of the Earth-like planet (Case No.6)}
\begin{tabular*}{170mm}{@{\extracolsep{\fill}}cccccccc} \hline
   Collision Time & $\mathrm{ID}_{\mathrm{imp}}$ & $\mathrm{ID}_{\mathrm{tar}}$ & $M_{\mathrm{imp}}\, [M_{\oplus}]$ & $M_{\mathrm{tar}}\, [M_{\oplus}]$ & $M_{\mathrm{merge}}\, [M_{\oplus}]$ & $a\, [\mathrm{AU}]$ & $M_{\mathrm{atm}}$\\
   \hline \hline
   0.057  Myr & 15 & 17 & 0.100 & 0.107 & 0.207 & 1.077 & $\gg2.0\,\mathrm{wt\%}$ \\
   0.451 Myr & 18 & 17 & 0.111 & 0.207 & 0.319 & 1.152 & $\gg2.0\,\mathrm{wt\%}$ \\
   0.797 Myr & 19 & 17 & 0.115 & 0.319 & 0.434 & 1.077 & $\gg2.0\,\mathrm{wt\%}$ \\
   \hline
    & & & & & & & $0.192\,\mathrm{wt\%}\, \, (M_{\mathrm{cylinder}})$ \\
   4.486 Myr & 8 & 10 & 0.079 & 0.166 & 0.245 & 0.697 & $0.105\,\mathrm{wt\%}\, \, (M_{\mathrm{mid}})$ \\
    & & & & & & & $0.0182\,\mathrm{wt\%}\, \, (M_{\mathrm{torus}})$ \\
   \hline
   8.6 Myr & 20 & 17 & 0.120 & 0.434 & 0.554 & 1.148 & $\sim0$ \\
   17.7 Myr & 10 & 17 & 0.245 & 0.554 & 0.799 & 1.066 & $\sim0$ \\
   69.7 Myr & 5 & 17 & 0.072 & 0.799 & 0.871 & 1.085 & $\sim0$ \\
   \hline
\end{tabular*}
\end{center}
\tablecomments{Collision time, IDs of the impactor and target (the smaller-mass body is the impactor), masses of the impactor and target, mass of the merged body, post-impact semimajor axis, and atmospheric acquisition. For atmospheric mass, we show $M_{\mathrm{mid}}$ unless otherwise noted. We list only the giant impacts for the planet that ultimately formed near 1 AU (ID=17). Initial giant impacts result in significant hydrogen absorption, while later giant impacts occur after disk gas dissipation.}
\label{tab:collision timeline}
\end{table}

Ultimately, our simulations identified one case that produced both an Earth-like third planet and a system of four terrestrial planets (see Case No.~6 in Figure~\ref{fig:chemical_evolution_3D_No6}). In this case, the third planet has a mass of $0.871 \, M_{\oplus}$ and a core density deficit of $8.58^{+1.54}_{-0.44}$\%, and it forms at an orbital distance of 1.085 AU with an eccentricity of $e = 0.034$. While constraints on Earth's core hydrogen content are improving, multiple giant impacts remain essential to ``adjust'' the hydrogen inventory. Whether an Earth-like composition emerges depends largely on whether a hydrogen-rich protoplanet collides with a hydrogen-poor one, thereby balancing the hydrogen budget.

From the distribution of core density deficit of finally-formed planets, larger planetary mass tends to decrease core density deficit(Figure \ref{fig:distribution of core DD}). Planets around $1 M_{\oplus}$ are roughly within the range of the current core density deficit of the Earth. In this sense, formation of the Earth in the solar system is statistically consistent with our results. 

\begin{figure}[ht!]
\plotone{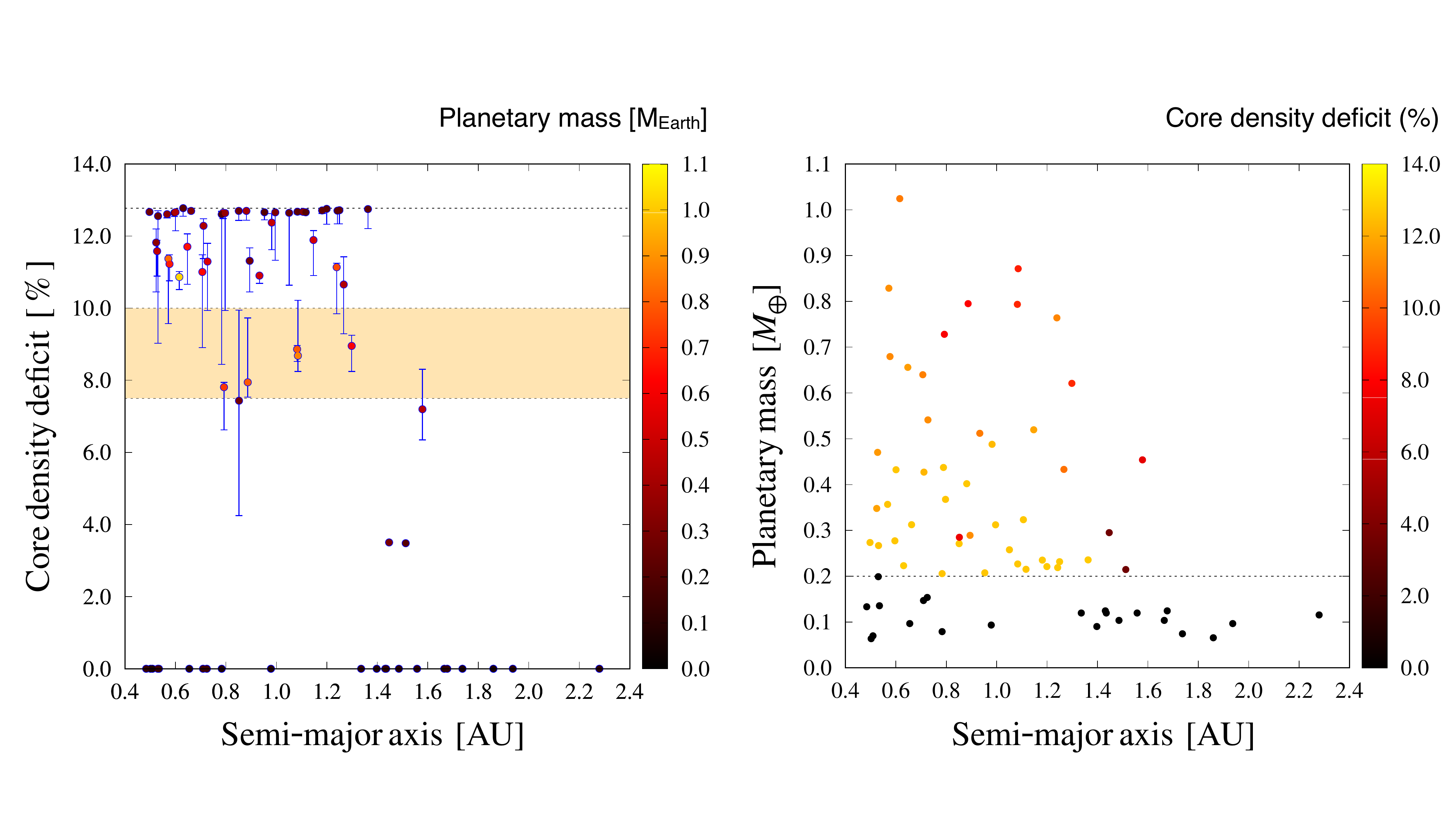}
\caption{A distribution of core density deficit for semi-major axis and planetary mass. The core density deficit is the results with an atmospheric mass of $M_{\mathrm{mid}}$. As the planetary mass increases, the core density deficit tends to decrease. On the other hand, there seems to be no apparent correlation between semi-major axis and core density deficit. }
\label{Figure:distribution_a_M_coreDD}
\end{figure}

Finally, we summarize the full dynamical and chemical evolution of the protoplanetary system. For Case No.6, the giant impact history of the planet forming near 1 AU is shown in Table~\ref{tab:collision timeline}, and the corresponding evolution of the orbital elements and core density deficit is shown in Figure~\ref{fig:chemical_evolution_3D_No6}. We note that this particular protoplanet (ID=17) initially accumulated excess hydrogen during early giant impacts in a dense gas disk. However, subsequent collisions—occurring after disk gas had largely dissipated—moderated its core density deficit to near-Earth values. These results highlight that initial giant impacts tend to over-enrich cores in hydrogen, and that late giant impacts involving hydrogen-poor protoplanets are critical to achieving an Earth-like outcome. Taken together with current geophysical constraints (7.5--10\% core density deficit), our simulations suggest a plausible multi-stage pathway for forming an Earth-like planet at 1 AU.

\subsection{A robustness of the chemical calculation}
\label{discussion:robustness of chemical calculation}

In the re-equilibration calculations, we do not consider any additional chemical variations induced directly by giant impacts themselves. However, the effects of strong shocks on the planetary mantle and core caused by giant impacts should be taken into account. \citet{Kurosaki2023} focused on chemical reactions between a hydrogen-rich atmosphere and rocky material vaporized by impact shocks, and calculated the resulting atmospheric compositions after giant impacts. They showed that a high mixing ratio of rocky vapor in the atmosphere can lead to an enhanced production of water. Some of the produced water can decompose back into hydrogen, potentially reducing the amount of hydrogen returned from the core to the atmosphere. Nevertheless, because the changes in the core density deficit due to re-equilibration are originally small, our main conclusions are not significantly affected. The final amount of water on the planet would increase; however, this remains highly uncertain given the contributions from other sources of water supply.

\begin{figure}[ht!]
\epsscale{1.0}
\plotone{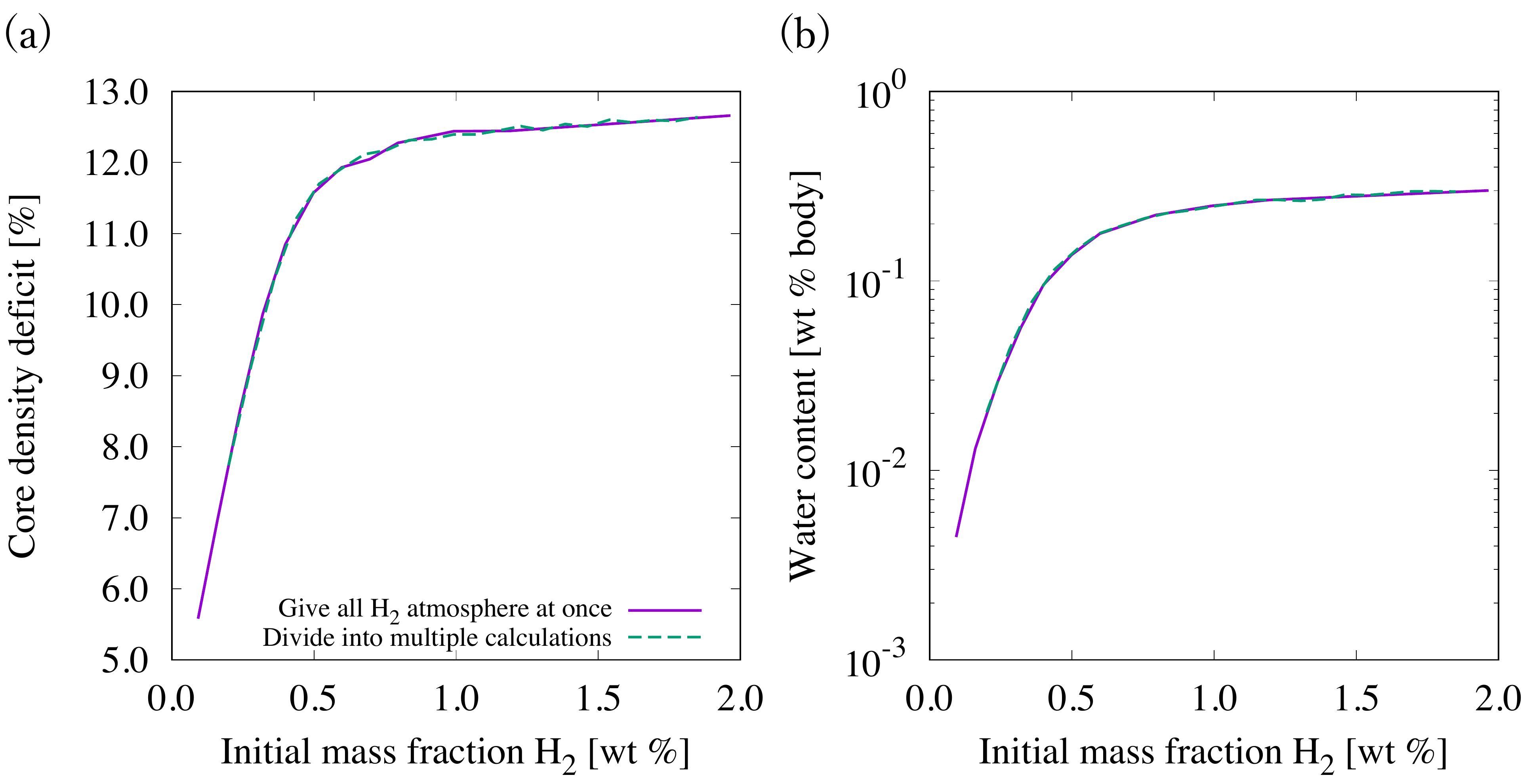}
\caption{Initial $\mathrm{H}_{2}$ mass fraction and (a) core density deficit and (b) water content, comparing cases where all hydrogen is supplied at once versus supplied incrementally. For example, to model a 0.5 wt\% hydrogen atmosphere, five equilibrium calculations were performed, each adding 0.1 wt\% hydrogen ($0.1 + 0.1 + 0.1 + 0.1 + 0.1 = 0.5$). The incremental addition better reflects physical constraints on atmosphere retention. Nevertheless, the differences between the two approaches are minor.
\label{fig:comparison}}
\end{figure}

Chemical equilibrium calculations account for the effect of atmospheric pressure at the planetary surface, which enhances the dissolution of atmospheric components into the magma ocean. This effect could potentially alter the results, especially when more disk gas surrounds the protoplanet than it can gravitationally bind. To assess this, we compared results obtained by supplying the atmosphere all at once (high initial atmospheric pressure) versus dividing it into increments with re-equilibration after each addition. Specifically, we modeled a protoplanet with a mass of 0.5\,$M_{\oplus}$, initially endowed with more atmosphere than it could stably retain (0.2 wt\%), using two approaches: (1) a single equilibrium calculation with the entire atmosphere and (2) sequential equilibrium calculations, adding 0.1 wt\% hydrogen at each step. We found no significant differences in the final chemical compositions between the two approaches (Figure~\ref{fig:comparison}). Although the surface pressure varied during the incremental calculations, its effect on the final composition was negligible.

\begin{figure}[ht!]
\plotone{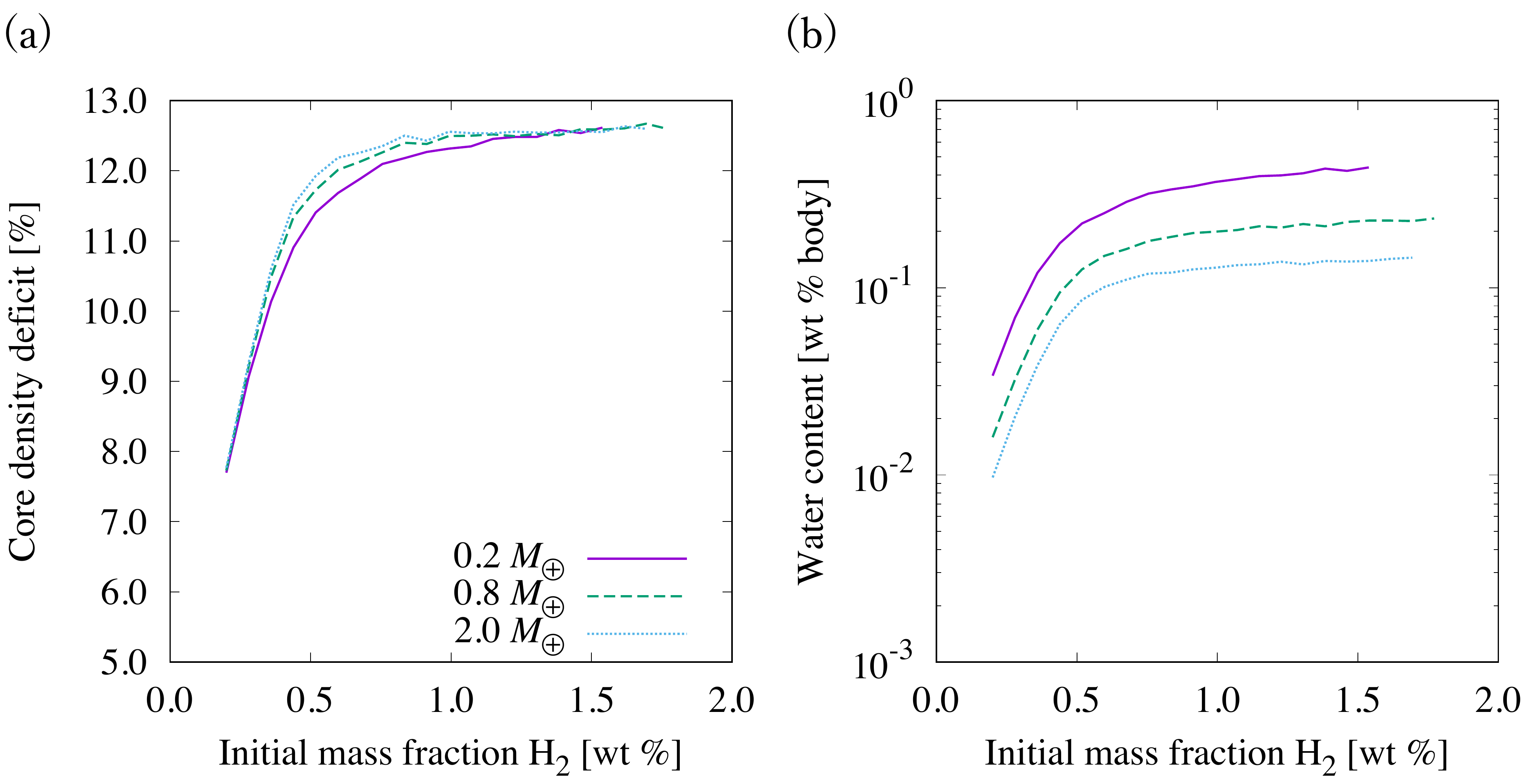}
\caption{Initial $\mathrm{H}_{2}$ mass fraction and (a) core density deficit and (b) water content as a function of protoplanet mass. Increasing protoplanet mass results in higher surface pressure, leading to greater hydrogen dissolution into the interior and less water production per unit mass. This calculation modifies only the protoplanet mass $M_{\mathrm{proto}}$ (and the corresponding initial $\mathrm{H}_{2}$ mass fraction range) relative to the models of \citet{young23}.
\label{fig:compare_mass}}
\end{figure}

To further investigate surface pressure effects, we conducted calculations where the protoplanet mass was varied as a parameter. The results show a slight variation in core density deficit and a moderate difference in water production (Figure~\ref{fig:compare_mass}). As protoplanet mass increases, surface pressure increases for a given atmospheric mass, leading to enhanced hydrogen dissolution into the interior and reduced water production per unit mass.

The surface pressure $P_{\mathrm{surface}}$ was calculated following \citet{young23}:
\begin{equation}
\left( \frac{P_{\mathrm{surface}}}{1 \, \mathrm{bar}} \right) = 1.2 \times 10^6 \left(\frac{M_{\mathrm{atm}}}{M_{\mathrm{planet}}}\right)\left(\frac{M_{\mathrm{planet}}}{M_{\oplus}}\right)^{2/3},
\end{equation}
where $M_{\mathrm{atm}}$ is the atmospheric mass and $M_{\mathrm{planet}}$ is the protoplanet mass. For a constant atmospheric-to-planet mass ratio, larger protoplanets produce higher surface pressures, promoting more efficient hydrogen dissolution into the interior and thus reducing water production per unit mass.

\subsection{Accretion of disk gas and atmospheric gain}

Protoplanets acquire their primordial atmospheres by gravitationally capturing gas from the surrounding protoplanetary disk. In our calculations, we assume that a protoplanet captures all disk gas within $1\,r_{\mathrm{Hill, s}}$ radially and set three models of atmospheric mass depending on how the disk gas is gathered in the vertical direction (Equation \ref{eq:Atm torus} and \ref{eq:Atm cylinder}). The ``cylindrical'' model is an upper limit and the ``toroidal'' model is a lower limit in our assumption.

Comparing gas capture estimates between the cylindrical and toroidal models, we find:
\begin{equation}
\frac{M_{\mathrm{torus}}}{M_{\mathrm{cylinder}}} 
= \frac{\pi \,r_{\mathrm{Hill, s}} \,\rho^{\mathrm{min}}_{\mathrm{gas}}}{2\,\Sigma^{\mathrm{min}}_{\mathrm{gas}}}
\sim 0.11 \left( \frac{M_{\mathrm{proto}}}{0.5 \, M_{\oplus}} \right)^{\frac{1}{3}} \left( \frac{a}{1 \, \mathrm{AU}} \right)^{-\frac{1}{4}}.
\end{equation}
This implies that the atmospheric mass models have indeterminacy by about one order of magnitude.

Given the disk conditions assumed in this study ($\Sigma_{\mathrm{gas}}(0) = 0.01 \, \Sigma^{\mathrm{min}}_{\mathrm{gas}}$ and $\tau_{\mathrm{diss}} = 10^{6}$ yr), initial giant impacts occur by $t \simeq 1\,\mathrm{Myr}$ at the latest. At this time, the remaining disk gas fraction is:
\begin{equation}
\kappa_{t = 1 \, \mathrm{Myr}}
= 0.01 \,\exp\left( -\frac{1 \,\mathrm{Myr}}{1 \,\mathrm{Myr}} \right)
\simeq 0.0037 \quad (\simeq 2.56\,\kappa_{\mathrm{torus}}).
\end{equation}
Thus, even if we adopt the lower limit atmospheric mass $M_{\mathrm{torus}}$, protoplanets still acquire more hydrogen than needed to reproduce Earth-like compositions during the initial giant impacts. Furthermore, once hydrogen saturation is achieved in the iron core, additional hydrogen intake has only a limited impact on the final core density deficit, preserving the system's overall characteristics.

On the other hand, protoplanetary systems have some giant impacts also in 1.9 - 4.2 Myr, result in significantly different atmospheric mass acquisition. Throughout the whole evolution of the system, this difference is moderated to some extent by multiple giant impacts, including initial and late ones. The distribution of the finally-formed planets shows a difference of about 2 \% in the core density deficit for planets growing to about one earth mass by lower and upper limits of atmospheric masses (See Figure \ref{fig:distribution of core DD}). In this sense, planet formation by multiple giant impacts makes the final chemical composition robust.

In addition to the Hill radius, the Bondi radius provides another scale relevant for gas accretion, defined by equating a planet's escape velocity with the thermal velocity of the surrounding gas. For protoplanets in the mass range $0.2$–$1\,M_{\oplus}$, the Bondi radius can be roughly an order of magnitude smaller than $r_{\mathrm{Hill}}$. Thus, the effective capture radius $\Delta r$ could vary by up to a factor of 10, modifying the gas accretion rate accordingly. However, once a hydrogen-rich atmosphere exceeds the core's saturation threshold (Figure~\ref{fig:H2-core_DD}), further increases in atmospheric hydrogen do not significantly affect the equilibrium composition or the core density deficit.

Even in an extreme case where a protoplanet captures gas extending out to $10\,r_{\mathrm{Hill}}$—intruding into neighboring protoplanets' gravitational domains—the atmosphere gained would still exceed the amount needed for hydrogen saturation. Therefore, despite large uncertainties in the capture radius (Hill vs.\ Bondi) and simplifications in our capture model (cylindrical vs.\ toroidal geometry), the key chemical outcomes remain qualitatively robust. Once hydrogen saturation is reached, additional gas capture has only a minor effect on the final chemical composition of the protoplanet.

\subsection{Atmospheric blowout}

In our $N$-body calculations, there is an interval of approximately $10^7$ years between the clustered initial giant impacts and the late giant impacts that occur after most of the disk gas has dissipated. According to \citet{hamano13}, Earth's surface solidified within a few million years, allowing oceans to form, whereas Venus may have remained molten for up to $\sim 10^{8}$ years and subsequently lost its water over this extended period. Therefore, protoplanets located beyond Earth's orbit would have had sufficient time between the first and second giant impacts to solidify their magma oceans and develop transient surface oceans.

When a protoplanet with a liquid ocean undergoes a giant impact, the impact energy vaporizes surface water; the resulting water vapor explosion can blow off much of the remaining hydrogen atmosphere if it is sufficiently thin \citep{genda05}. In this scenario, the water vapor produced by the impact is retained and eventually contributes to the formation of permanent oceans on Earth-like planets. Thus, any excess hydrogen acquired during the initial giant impacts can be effectively stripped away, resetting the atmospheric composition.

By contrast, if hydrogen remains bound to the protoplanet after a giant impact—either because oceans are absent or because processes such as atmospheric mixing help retain hydrogen—then the hydrogen initially incorporated into the core during the initial giant impacts may remain locked within the planet, limiting its subsequent release into the atmosphere. Recent numerical simulations suggest that oblique giant impacts often remove only a modest fraction of a planet's hydrogen atmosphere \citep[e.g.,][]{Kegerreis2020}, and that planets can retain over $60\%$ of their pre-impact atmospheres \citep{Matsumoto2021}. Such partial atmospheric retention can, in turn, influence the final core density deficit and water inventory, reflecting a complex interplay between impact geometry, planetary rotation, and disk gas availability.

\subsection{Final number of protoplanets} \label{sec:final_number_planet}

\citet{kominami02} showed that even after complete gas dissipation, once protoplanets achieve sufficiently large orbital separations ($\gtrsim 20\,r_{\mathrm{Hill}}$), their orbits can remain stable over long timescales. In their simulations, which extended up to about $10^7$ years, they found that more protoplanets tended to survive than the four rocky planets observed in the present solar system. We confirmed this trend in our own longer-duration calculations (up to $10^8$ years). Although 5 out of 12 simulations resulted in systems with four or fewer planets, some cases produced more than twice as many planets as in the present-day solar system. Interestingly, the presence of disk gas appears to suppress giant impacts among protoplanets, contrasting with the gas-free calculations by \citet{kokubo06}, whose average final planet count differs significantly from our results.

Specifically, \citet{kokubo06} explored a range of models without gas drag and found final planet numbers of $\langle N_{\mathrm{proto}} \rangle_{\mathrm{min}} = 2.3$ and $\langle N_{\mathrm{proto}} \rangle_{\mathrm{max}} = 4.4$ over 10 different setups. They performed 20 runs for each model, varying the initial azimuthal distributions of protoplanets. Even in a scenario closely matching our setup—with $\Delta a = 8\,r_{\mathrm{Hill}}$, $a_{\mathrm{min}}=0.5\,\mathrm{AU}$, $a_{\mathrm{max}}=1.5\,\mathrm{AU}$, $N_{\mathrm{proto}}=23$, and $M_{\mathrm{total}}=2.40\,M_{\oplus}$—their average outcome was $\langle N_{\mathrm{proto}} \rangle = 3.5 \pm 0.8$, with an accretion timescale $T_{\mathrm{acc}}=0.87 \pm 0.40 \times 10^{8}\,\mathrm{yr}$ (where $T_{\mathrm{acc}}$ is defined as the interval between the first and last collisions).

They also showed that the total mass of protoplanets $M_{\mathrm{total}}$ plays a key role in determining the final number of planets, following the relation:
\begin{equation}
\langle N_{\mathrm{proto}} \rangle \simeq 3.5 
\left( \frac{M_{\mathrm{total}}}{2\, M_{\oplus}} \right)^{-0.15}.
\label{eq:number of planet for total mass}
\end{equation}
In our study, the total mass was $M_{\mathrm{total}} \simeq 1.88\,M_{\oplus}$; however, the weak exponent ($-0.15$) implies that this difference is unlikely to fully explain the discrepancy in final planet counts.

\citet{kominami02} suggested that further reductions in the number of surviving protoplanets may require additional perturbations, such as those induced by gas giants, which can eject or merge nearby protoplanets through strong gravitational interactions. In our study, giant impacts occurring after complete gas dissipation do not significantly reduce planet numbers; thus, incorporating the dynamics of gas giants may be necessary to fully reproduce the architecture of the solar system.

In our calculations, the mean final number of planets remains $\langle N_{\mathrm{planet}} \rangle = 5.75$ at $t = 10^8$ years (averaged over 12 runs; see Table~\ref{tab:inclination N-body}). This result supports the notion that additional large-scale dynamical perturbations—beyond those considered in this study—are likely necessary to reproduce a system with a small number of rocky planets, as seen in the solar system.

We terminated our orbital evolution calculations at $10^8$ years. However, it is possible to estimate the orbital crossing timescale at that epoch. \citet{Yoshinaga1999} investigated the dynamical instability of protoplanetary systems with nonzero eccentricities and inclinations. Although they did not consider gas drag, by $10^8$ years in our simulations, the disk gas has almost completely dissipated. We estimated the instability timescale for the systems listed in Table~\ref{tab:inclination N-body}, focusing on the smallest orbital separations $\Delta a_{\mathrm{min}}$. The results are shown in Table~\ref{tab:instability estimate}. These estimates from $\Delta a_{\mathrm{min}}$ provide lower limits, since they are based on the most closely packed protoplanet pairs in each system \citep{Wu2019}.

The instability timescale depends strongly on the initial random velocities (eccentricities and inclinations) and orbital separations $\Delta a$. For some cases, further giant impacts may occur and reduce the final number of planets. However, even if additional giant impacts occur, they would not significantly alter the main conclusions of this study. After disk gas dissipation, the precise timing of giant impacts is less critical. Indeed, further giant impacts would likely help to regulate the hydrogen inventory in protoplanetary cores, increasing the probability of forming Earth-like planets. 

\citet{Yoshinaga1999} assumed that the eccentricity dispersion is twice as large as
the inclination dispersion ($\langle \tilde{e}^{2} \rangle^{1/2}$ = 2 $\langle \tilde{i}^{2} \rangle^{1/2}$) and set initial conditions. In our calculations, about half of the cases roughly match this assumption. Some studies provide
estimations of the orbital stability without this assumption \citep{pu2015, Wang2019}. A more detailed estimation or extended integrations would allow for a more accurate assessment of the effects of gas drag and other dynamical mechanisms on the final number of planets.

\begin{table}[h]
\begin{center}
\caption{Instability estimates for protoplanet systems at $t_{\mathrm{end}} =10^8$ years (corresponding to Table~\ref{tab:inclination N-body})}
\begin{tabular*}{170mm}{@{\extracolsep{\fill}}cccccc} \hline
   Case & Final $N_{\mathrm{proto}}$ & $\langle \tilde{e}^{2} \rangle^{1/2}$ & $\langle \tilde{i}^{2} \rangle^{1/2}$ & $\Delta a_{\mathrm{min}}\, [r_{\mathrm{Hill}}]$ & $\log\, T_{\mathrm{inst}}$ \\
   \hline \hline
   No. 1 & 4 & 28.6 & 23.8 & 10.0 & 5.14 \\
   No. 2 & 4 & 13.4 & 21.3 & 37.1 & 18.6 \\
   No. 3 & 5 & 15.2 & 10.1 & 9.25 & 4.75 \\
   No. 4 & 9 & 1.85 & 1.08 & 12.7 & 7.78 \\
   No. 5 & 9 & 2.16 & 1.25 & 13.8 & 8.47 \\
   No. 6 & 4 & 5.98 & 2.60 & 33.7 & 16.9 \\
   No. 7 & 6 & 4.44 & 2.40 & 20.6 & 10.4 \\
   No. 8 & 6 & 3.15 & 1.95 & 14.5 & 8.33 \\
   No. 9 & 9 & 0.568 & 0.877& 11.7 & 7.67 \\
   No. 10 & 6 & 5.76 & 2.54 & 21.1 & 10.6 \\
   No. 11 & 3 & 6.00 & 19.1 & 49.5 & 24.7 \\
   No. 12 & 4 & 26.9 & 13.9 & 36.6 & 18.3 \\
   \hline
\end{tabular*}
\end{center}
\tablecomments{Final number of protoplanets at $t=10^8$ years, RMS of scaled eccentricity $\tilde{e}$ and inclination $\tilde{i}$ (where $\tilde{e}=e/h$, $\tilde{i}=i/h$ and $h=r_{\mathrm{Hill}}/a$), minimum orbital separation $\Delta a_{\mathrm{min}}$, and estimated instability timescale $\log\,T_{\mathrm{inst}}$. Instability times are extrapolated based on \citet{Yoshinaga1999}, using the closest matching $\langle \tilde{e}^{2} \rangle^{1/2}$ value. In \citet{Yoshinaga1999}, the relationship between $\langle \tilde{e}^{2} \rangle^{1/2}$ and $\langle \tilde{i}^{2} \rangle^{1/2}$ is assumed. These estimates suggest possible further giant impacts, but the overall conclusions of this study remain unaffected.}
\label{tab:instability estimate}
\end{table}

\subsection{Formation of Venus}

Among the rocky planets, Venus shares a mass similar to that of Earth ($\sim 0.815 \, M_{\oplus}$) and likely underwent a comparable evolutionary trajectory under the assumptions made in this study (i.e., protoplanets with masses exceeding $0.2 \, M_{\oplus}$ possess atmospheres \citet{young23}). 

Our results suggest that Venus may have a core density deficit similar to that of Earth (See Figure \ref{Figure:distribution_a_M_coreDD}). Core density deficit and water content have no apparent correlation with orbital length radius. Core density deficit tend to be lower in the Earth and Venus orbits than at either end of the rocky planet region due to more giant impacts. In our calculations, the orbits of Earth and Venus seem too close together to distinguish between the two planets. When we focus on the planetary mass corelation, Venus may have acquired slightly more water than Earth during its formation (Figure \ref{fig:distribution of core DD}). 

A key determinant distinguishing the environments of the present-day Earth and Venus is the disparity in received solar radiation due to their varying distances from the Sun, which may have led to significant differences in the solidification times of their respective magma oceans \citep{hamano13}. Although this study does not incorporate the thermal evolution of protoplanets, including the effects of central stellar radiation, future investigations into atmospheric evolution are essential for understanding planet formation in similar environments, such as Earth and Venus.

We note that constraints by other chemical species also exist. In this study, which focuses on the chemical evolution associated with nebular gas, the presence of noble gases, especially Ne and Ar can be a tracer of disk gas. Although it is successful in explaining the difference in noble gas content between the Earth and Venus \citep{genda05}, difficulties remain in explaining the observed ratio of Ne to Ar. D/H ratio is another important tracer of nebular hydrogen. \citet{ikoma06} were forced to argue that the close similarity in D/H in Earth's water to D/H in meteorites is an accident. They had to argue that Earth lost just the right amount of H to raise the D/H ratio of Earth's water to make it look like the water in meteorites. It is possible but not as simple as just using materials resembling meteorites as the source.

\subsection{For the case of pebble accretion}

This study followed the evolution of protoplanets into planets based on the classical theory of planet formation. Recently, pebble accretion has been proposed as an efficient mechanism for building solid cores \citep{lambrechts12}, and scenarios incorporating pebble accretion have been advanced for the formation of terrestrial planets \citep{johansen21}. In contrast to the traditional model—which relies on collisions among kilometer-scale planetesimals—pebble accretion involves the gradual accretion of meter-sized or smaller particles, often referred to as pebbles. Shifting to a pebble-accretion framework thus requires rethinking the conditions of the protoplanetary disk, including the timing of gas dissipation and the gas surface density during protoplanet growth.

Although the present work does not focus on the origin of Earth's water, pebble accretion remains an important alternative scenario to the classical model adopted here. In particular, pebble accretion assumes that rocky planets form in a gas-rich environment, which can result in protoplanets being systematically more hydrogen-rich. This contrasts with our scenario, where the formation of an Earth-like planet requires at least one giant impact involving a hydrogen-poor protoplanet to dilute the excess hydrogen retained in hydrogen-rich cores. If all protoplanets are hydrogen-rich under pebble accretion, such ``balancing'' collisions may not occur, suggesting a potential incompatibility between these two models.

These estimates are just hypotheses, and further investigations into pebble accretion—especially concerning how much hydrogen can be acquired, redistributed, or lost during giant impacts—are needed. Integrating pebble accretion into $N$-body simulations and chemical equilibrium models, along with more detailed disk and impact physics, would clarify whether and how Earth-like planets could emerge under this alternative growth regime.

\section{Conclusion} \label{sec:conclusion}

In this study, we aimed to elucidate how a terrestrial planet with an Earth-like composition—featuring a hydrogen-rich primordial atmosphere, a magma ocean, and an iron core—could emerge from the protoplanetary stage. To this end, we combined an $N$-body planet formation model with a chemical equilibrium framework adapted from \citet{young23}, focusing on how multiple giant impacts affect the distribution of hydrogen (and, by extension, water) in forming planets.

Our results indicate that protoplanets can acquire a significant amount of hydrogen during early giant impacts in a gas-rich disk environment, raising the core density deficit well above Earth-like values. Once the disk gas dissipates, however, subsequent giant impacts occurring in a gas-poor setting allow some of this hydrogen to re-enter the atmosphere, contributing to water production. Crucially, late collisions between hydrogen-rich and hydrogen-poor protoplanets can dilute the core's hydrogen content, thereby reducing the density deficit to Earth-like levels.

A core density deficit near Earth's observed 7.5 - 10\% emerges when at least one protoplanet experiences early hydrogen over-enrichment followed by a late collision with a hydrogen-poor partner. Although our calculations account for water formation from hydrogen returning to the atmosphere, we do not treat the final water inventory as a stringent constraint, given that Earth likely received water from multiple exogenous sources. Instead, the iron core's hydrogen budget serves as our principal diagnostic, aligning with Earth's observed density deficit.

By coupling $N$-body orbital evolution and chemical equilibrium calculations, we highlight a strong feedback between dynamical and chemical processes during rocky planet formation: dynamical collisions dictate hydrogen retention or removal at each stage, while the evolving hydrogen inventory influences subsequent collision outcomes. This interplay underscores the complexity of terrestrial planet assembly. Our scenario does not exactly reproduce the solar system's current mass distribution; rather, it demonstrates that at least one protoplanet near 1~AU can acquire an Earth-like chemical state under realistic disk dissipation and impact sequencing conditions.

Although we adopt simplified prescriptions for gas capture (e.g., Hill-sphere approximations), this approach clarifies how hydrogen-saturated cores can evolve into Earth-like configurations through repeated impacts and core mixing. As future work refines material properties, disk evolution timescales, and impact physics, this integrated dynamical–chemical framework offers a promising pathway for understanding rocky planet origins—both within our solar system and in extrasolar systems.

\begin{acknowledgments}
We thank anonymous reviewers for their constructive comments, which led us to greatly improve this paper. HM also thanks Kumano Dormitory community at Kyoto University for their generous financial and living assistance. This work was supported by JSPS KAKENHI Grant Number 21H04512.
\end{acknowledgments}

\appendix

\section{Calculations for the Zero Inclination Cases}

\subsection{Results of $N$-body Simulation}

Here, we summarize the results of $N$-body calculations and the chemical evolution of protoplanets for the case where protoplanets move on a coplanar surface. As a first step, we set the initial orbital inclinations of the protoplanets to zero. This approach allows us to systematically explore a broad parameter space and identify key trends in planet formation under various gas disk conditions. By analyzing the results obtained in this simplified setting, we can efficiently determine which parameter ranges are most relevant for more detailed simulations with nonzero inclination. This strategy ensures that subsequent inclined-orbit simulations are focused on the most promising conditions for understanding realistic planetary formation scenarios. We performed numerous simulations under various conditions, with the primary parameters being $\Sigma_{\mathrm{gas}}(0)$ and $\tau_{\mathrm{diss}}$.

\begin{figure}[ht!]
\plotone{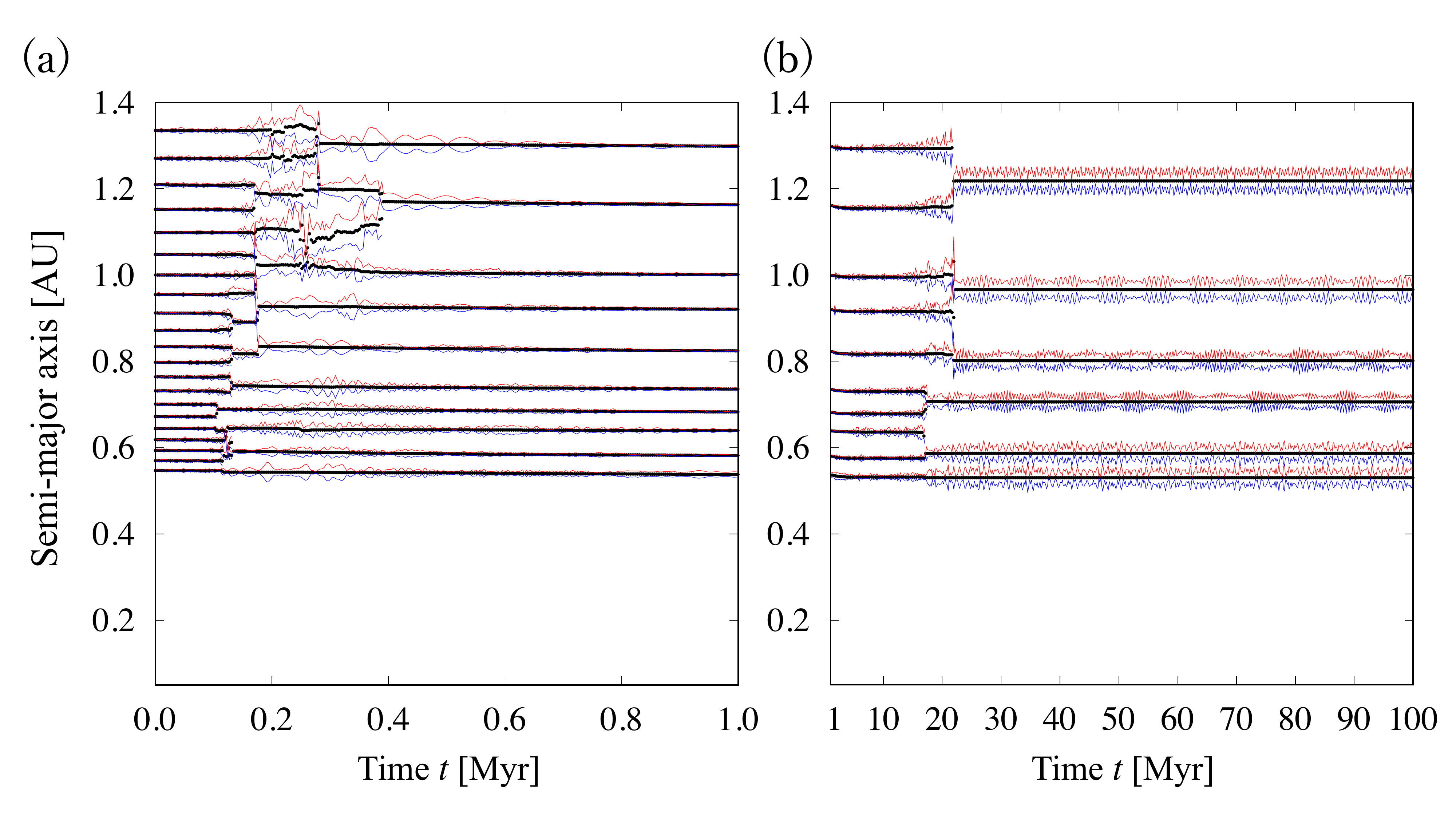}
\caption{Orbital evolution of the protoplanets for (a) $0 \leq t \leq 1.0 \, \mathrm{Myr}$ and (b) $1.0 \leq t \leq 100 \, \mathrm{Myr}$, assuming $\Sigma_{\mathrm{gas}}(0) = 0.01\, \Sigma_{\mathrm{gas}}^{\mathrm{min}}$ and $\tau_{\mathrm{diss}} = 10^6 \, \mathrm{yr}$. The black dots represent the semimajor axis, while the blue and red lines indicate perihelion and aphelion distances, respectively. Giant impacts cluster at $t \simeq 0.1 \, \mathrm{Myr}$ (first stage) and $t \simeq 20 \, \mathrm{Myr}$ (second stage).}
\label{fig:orbit_time_profile_2}
\end{figure}

\begin{figure}[ht!]
\plotone{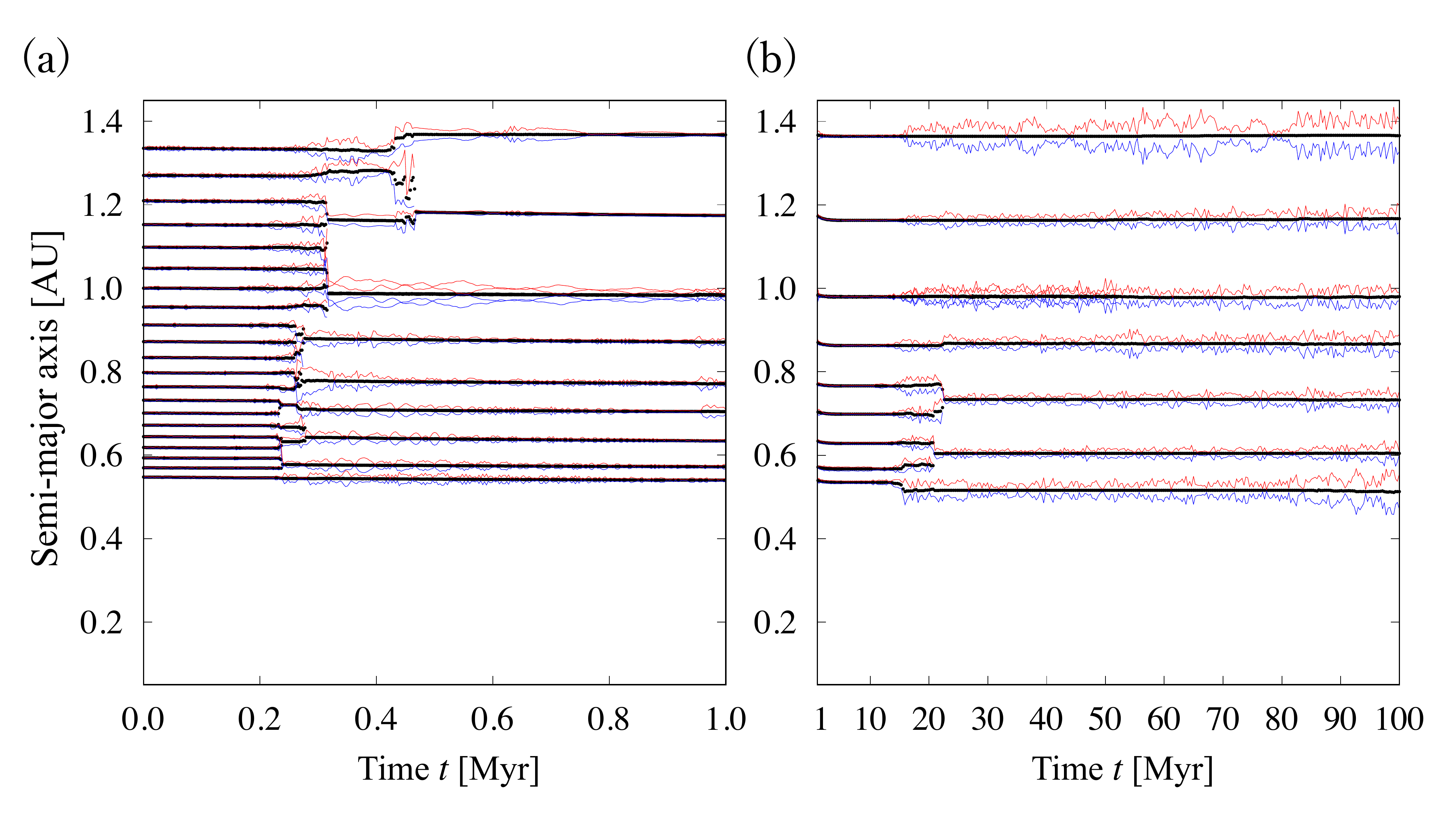}
\plotone{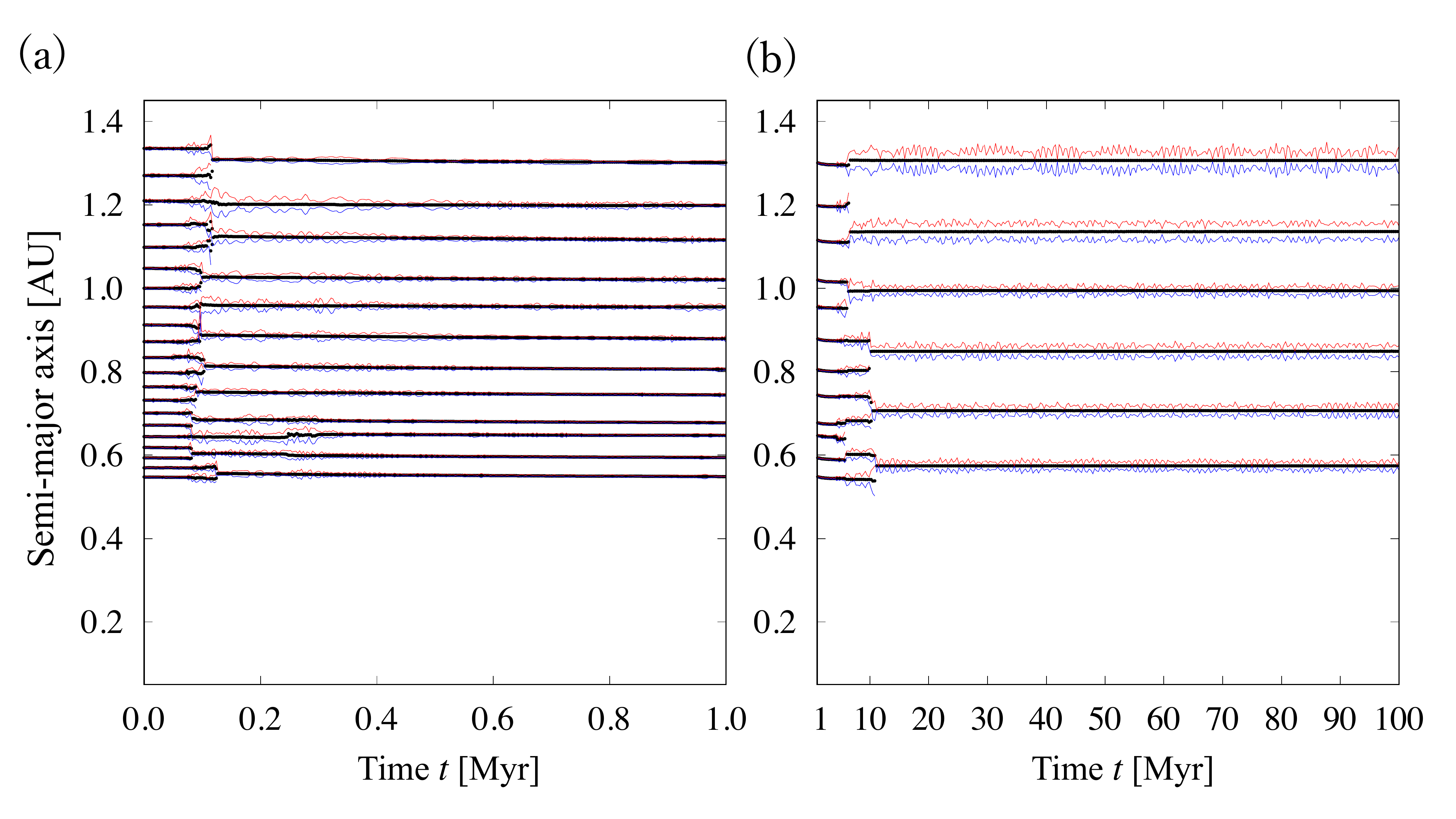}
\caption{Orbital evolution of the protoplanets for two additional runs using the same parameters as Figure~\ref{fig:orbit_time_profile_2}, but with different random initial placements of protoplanets. (a)--(b) show the system's evolution for $t\leq 1\,\mathrm{Myr}$ and $1\leq t\leq 100\,\mathrm{Myr}$, respectively. Random differences in initial azimuthal positions affect the timing and extent of the second giant impact stage.}
\label{fig:orbit_time_profile_3}
\end{figure}

In cases with $\Sigma_{\mathrm{gas}}(0) \sim 1.0\,\Sigma_{\mathrm{gas}}^{\mathrm{min}}$, gas dissipation proceeds too slowly, and the disk remains dense. As a result, protoplanets lose substantial angular momentum via gas drag; when $\tau_{\mathrm{diss}} \gtrsim 2\times10^5$ years, they migrate inward and accrete onto the star without undergoing mutual collisions. This infall behavior can be compared with that associated with so-called Type I migration \citep{Tanaka2002}.

On the other hand, in cases with more tenuous disks ($0.01$--$0.1\,\Sigma_{\mathrm{gas}}^{\mathrm{min}}$), giant impacts occur before the protoplanets spiral into the star. Figure~\ref{fig:orbit_time_profile_2} illustrates a typical run with $\Sigma_{\mathrm{gas}}(0)=0.01\,\Sigma_{\mathrm{gas}}^{\mathrm{min}}$ and $\tau_{\mathrm{diss}}=10^6$ years. The system undergoes a ``first giant impact'' stage ($t\sim 0.1$--$0.2$ Myr), followed by a quiescent interval lasting $\gtrsim 10^7$ years, and then a ``second giant impact'' stage.

By examining the evolution of perihelion and aphelion distances, we find that initially excited orbits are partially damped by gas drag up to about $10^{7}$ years. Once sufficient gas has dissipated, gravitational interactions destabilize the protoplanetary orbits again, triggering the second giant impact stage.

Because of the chaotic nature of $N$-body dynamics, random variations in the initial placement of protoplanets can significantly affect the long-term timeline of collisions. Figure~\ref{fig:orbit_time_profile_3} compares two additional runs with identical disk parameters but different random initial azimuthal distributions. In both cases, a two-stage giant impact sequence emerges, although the timing of the second impact stage can vary. Some outer-orbit protoplanets may avoid collisions beyond $10^8$ years if their orbits remain sufficiently isolated.

\subsection{Chemical Evolution of Protoplanets}

\begin{figure}[ht!]
\plotone{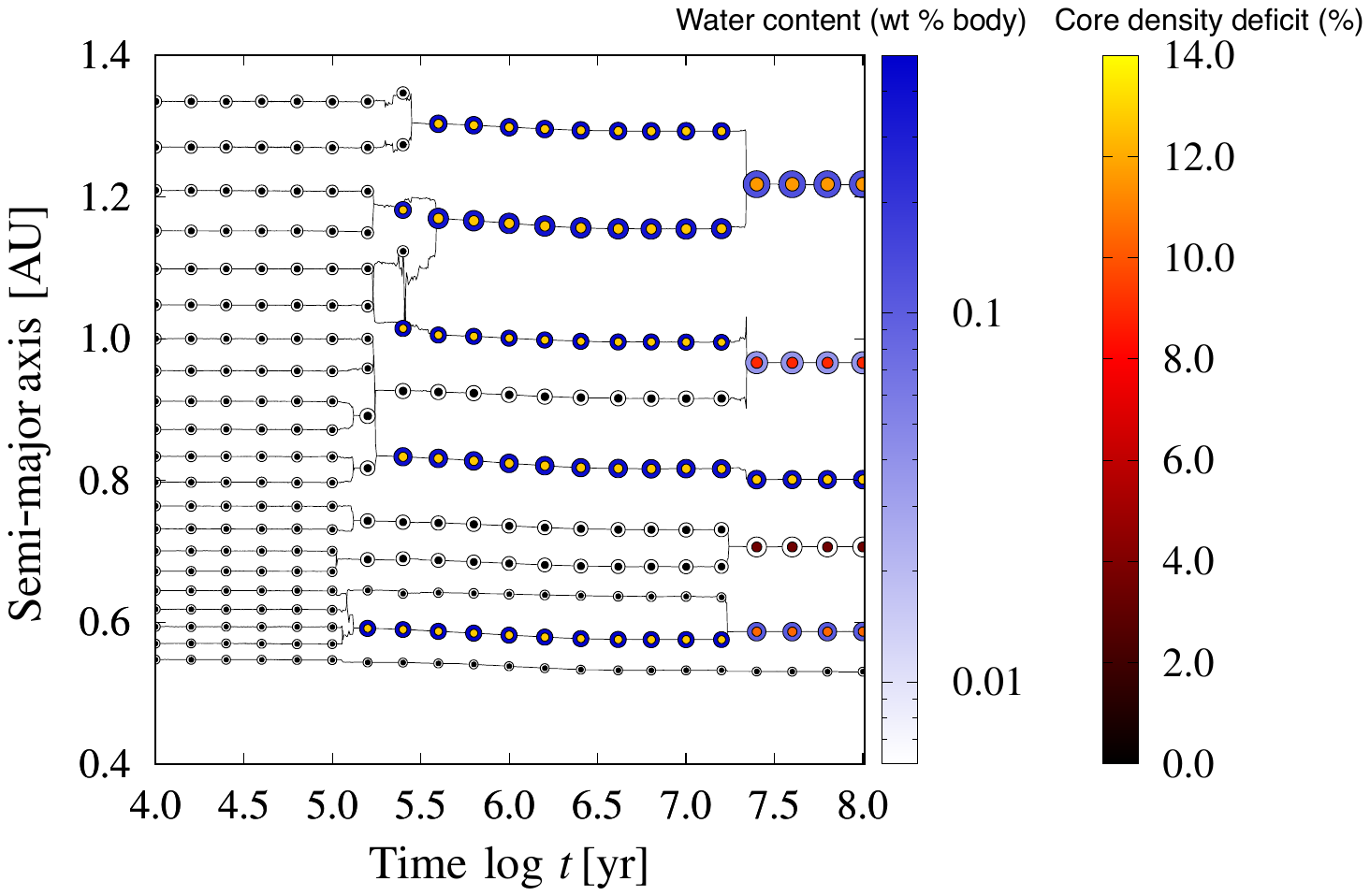}
\caption{Orbital evolution of the protoplanets and changes in core density deficit and water content for $\Sigma_{\mathrm{gas}}(0) = 0.01 \, \Sigma_{\mathrm{gas}}^{\mathrm{min}}$ and $\tau_{\mathrm{diss}} = 10^{6}$ yr. Circle colors denote the protoplanets' core density deficits and water contents. Initially, several protoplanets gain excess hydrogen in their cores (first giant impact stage), but subsequent collisions can moderate the core density deficit.}
\label{fig:chemical_evolution_2D_No1}
\end{figure}

We performed chemical equilibrium calculations for the protoplanets to track how the iron core density deficit evolves as they accumulate and lose hydrogen. Figure~\ref{fig:chemical_evolution_2D_No1} shows a representative case, corresponding to the same simulation presented in Figure~\ref{fig:orbit_time_profile_2}. During the first giant impact stage, five protoplanets acquire hydrogen-rich cores; some subsequently undergo further collisions that reduce their core density deficits.

In this example, a planet with a moderate core density deficit eventually forms near 1~AU. Four late-stage giant impacts occur, each with different characteristics: collisions between two hydrogen-rich cores slightly decrease the core hydrogen content; collisions between hydrogen-rich and hydrogen-poor cores yield a more moderate core density deficit, potentially producing Earth-like compositions; and collisions between two hydrogen-poor cores generally do not significantly alter the core composition.

Across a broad range of parameter space, we find that protoplanets in the first giant impact stage typically capture more atmosphere than is needed to reproduce Earth's composition, unless gas dissipation is extremely rapid ($\tau_{\mathrm{diss}} \lesssim 0.1$ Myr). This outcome highlights the importance of multiple late-stage giant impacts to adjust the hydrogen inventory, thereby increasing the likelihood of forming an Earth-like planet.

\subsection{Statistical Analysis and Parameter Sensitivity} \label{sec:statistic}

Our simulations consistently reveal a multi-stage growth pattern from protoplanets to planets: initially, giant impacts occur in a relatively dense disk, enabling substantial hydrogen capture into the iron core; in later impacts—after much of the disk gas has dissipated—protoplanets acquire little new atmosphere, but hydrogen already incorporated into the core can be partially released back into the atmosphere. Whether Earth-like compositions arise depends on the specific collision history within the network of protoplanets, reflecting the chaotic, sensitive nature of multi-body orbital dynamics \citep{kokubo06}.

\begin{table}[h]
\begin{center}
\caption{Results of statistical $N$-body simulations (after the first giant impact stage). Underlined rows correspond to runs highlighted in Figure~\ref{fig:N-body statistical calsulation}.}
\begin{tabular*}{150mm}{@{\extracolsep{\fill}}cccc} \hline
   $\Sigma_{\mathrm{gas}}(0)$ $[\Sigma^{\mathrm{min}}_{\mathrm{gas}}]$ & $t_{\mathrm{end}}$ [yr] & $N_{\mathrm{run}}$ & $\langle N_{\mathrm{proto}} \rangle \pm \sigma$  \\
   \hline \hline
   \uline{0.01} & $1.6 \times 10^{6}$ & 110 & $9.82 \pm 1.51$ \\
   0.01 & $3.2 \times 10^{6}$ & 50 & $9.42 \pm 1.77$ \\
   0.01 & $1.6 \times 10^{7}$ & 12 & $10.3 \pm 1.75$ \\
   \uline{0.02} & $1.6 \times 10^{6}$ & 150 & $10.3 \pm 1.46$ \\
   0.03 & $1.6 \times 10^{6}$ & 20 & $10.7 \pm 1.71$ \\
   \uline{0.03} & $3.2 \times 10^{6}$ & 85 & $10.7 \pm 1.48$ \\
   0.04 & $1.6 \times 10^{6}$ & 20 & $11.1 \pm 0.995$ \\
   \uline{0.04} & $3.2 \times 10^{6}$ & 85 & $10.6 \pm 1.79$ \\
   0.05 & $1.6 \times 10^{6}$ & 10 & $11.5 \pm 1.28$ \\
   \uline{0.05} & $4.8 \times 10^{6}$ & 98 & $10.6 \pm 1.67$ \\
   \uline{0.06} & $4.8 \times 10^{6}$ & 22 & $10.4 \pm 1.92$ \\
   \uline{0.07} & $4.8 \times 10^{6}$ & 21 & $10.6 \pm 1.84$ \\
   \uline{0.08} & $4.8 \times 10^{6}$ & 21 & $10.2 \pm 1.98$ \\
   \uline{0.09} & $4.8 \times 10^{6}$ & 21 & $10.7 \pm 2.25$ \\
   \uline{0.10} & $5.6 \times 10^{6}$ & 46 & $10.5 \pm 1.96$ \\
   \hline
\end{tabular*}
\end{center}
\tablecomments{Listed are the initial gas surface density $\Sigma_{\mathrm{gas}}(0)$, run duration $t_{\mathrm{end}}$, number of runs $N_{\mathrm{run}}$, and the average (with standard deviation) of protoplanets $\langle N_{\mathrm{proto}}\rangle$ at the end of the first giant impact stage. The initial 21-protoplanet systems typically end up with $\sim$10 protoplanets post-impact. Subsequent stages that further reduce the count closer to the modern solar system are not included.}
\label{tab:statistical N-body}
\end{table}

\begin{figure}[ht!]
\plotone{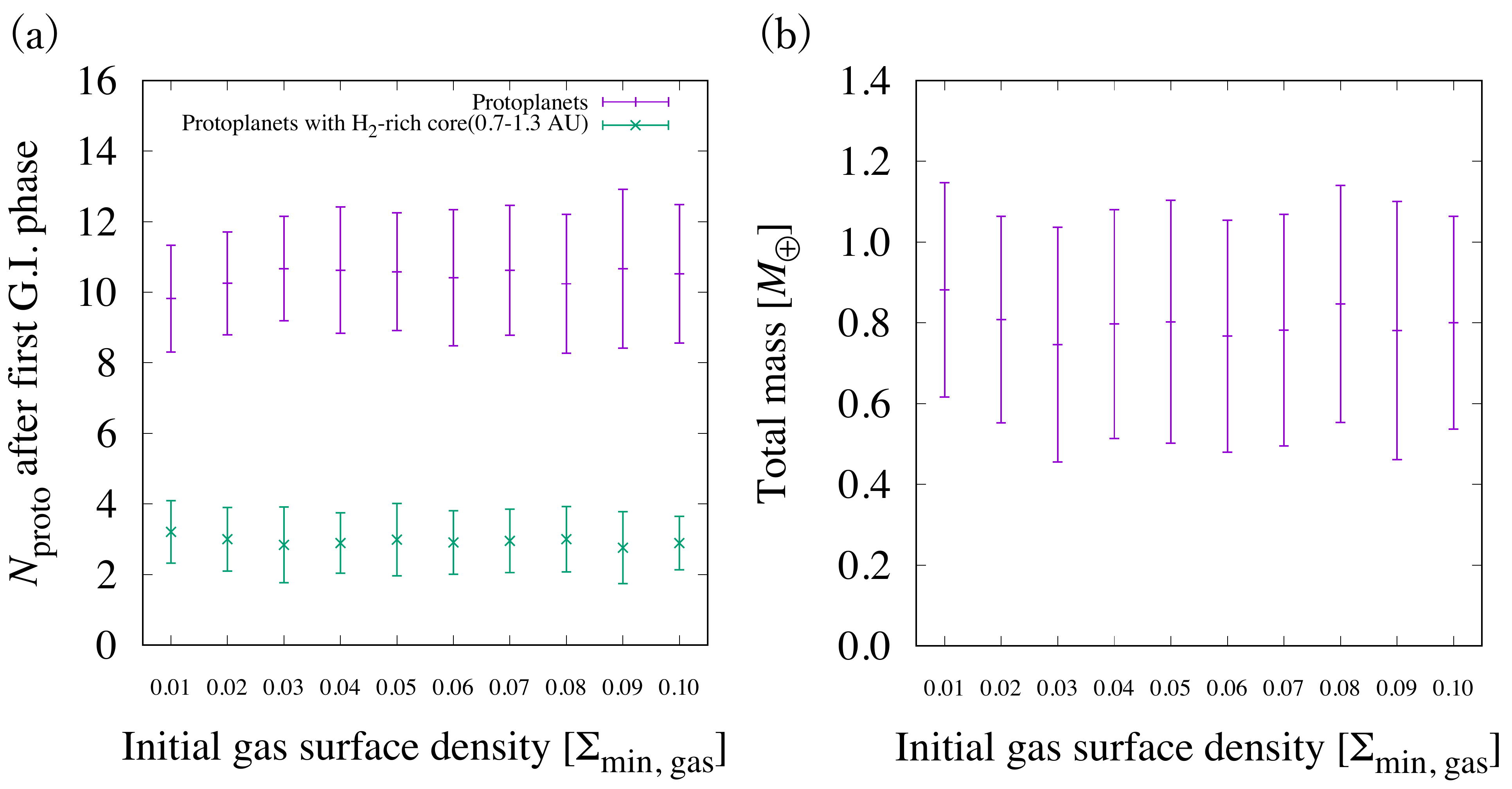}
\caption{(a) Number of protoplanets and hydrogen-rich-core protoplanets remaining after the first giant impact stage (0.7--1.3 AU). (b) Total mass of hydrogen-rich-core protoplanets, also 0.7--1.3 AU. The final number of protoplanets is almost independent of the initial gas surface density, suggesting that as long as the disk is sufficiently dense for collisions, the overall dynamical evolution—and hence the giant impact sequence—remains similar.}
\label{fig:N-body statistical calsulation}
\end{figure}

We carried out multiple runs with varied initial conditions to assess the robustness of our findings. Table~\ref{tab:statistical N-body} and Figure~\ref{fig:N-body statistical calsulation} summarize these statistical results. For most parameter choices, the first giant impact stage reduces the original 21 protoplanets to about half that number, typically leaving a total system mass of approximately one Earth mass contained in hydrogen-rich cores. Within the region from 0.7 to 1.3 AU, the total mass of protoplanets is often around $0.7\,M_{\oplus}$.

In particular, we find that the initial gas surface density and dissipation timescale have surprisingly little impact on the long-term dynamical outcomes, compared to the influence of random initial orbital placements. This insensitivity arises partly because the chemical equilibria saturate at high hydrogen contents. Once a protoplanet absorbs sufficient hydrogen, additional intake has diminishing effects on core composition. Thus, the overall collision outcomes are governed more by dynamical chaos than by fine-tuned variations in disk parameters.

\bibliography{maeda}{}

\begin{thebibliography}{}
\expandafter\ifx\csname natexlab\endcsname\relax\def\natexlab#1{#1}\fi
\providecommand{\url}[1]{\href{#1}{#1}}
\providecommand{\dodoi}[1]{doi:~\href{http://doi.org/#1}{\nolinkurl{#1}}}
\providecommand{\doeprint}[1]{\href{http://ascl.net/#1}{\nolinkurl{http://ascl.net/#1}}}
\providecommand{\doarXiv}[1]{\href{https://arxiv.org/abs/#1}{\nolinkurl{https://arxiv.org/abs/#1}}}

\bibitem[{Birch(1964)}]{birch1964}
Birch, F. 1964, Journal of Geophysical Research, 69, 4377.
\newblock \url{https://api.semanticscholar.org/CorpusID:129046042}

\bibitem[{{Bond} {et~al.}(2010){Bond}, {Lauretta}, \& {O'Brien}}]{bond2010}
{Bond}, J.~C., {Lauretta}, D.~S., \& {O'Brien}, D.~P. 2010, \icarus, 205, 321, \dodoi{10.1016/j.icarus.2009.07.037}

\bibitem[{{Chambers} \& {Wetherill}(1998)}]{chambers98}
{Chambers}, J.~E., \& {Wetherill}, G.~W. 1998, \icarus, 136, 304, \dodoi{10.1006/icar.1998.6007}

\bibitem[{{Fischer} {et~al.}(2017){Fischer}, {Campbell}, \& {Ciesla}}]{fischer2017}
{Fischer}, R.~A., {Campbell}, A.~J., \& {Ciesla}, F.~J. 2017, Earth and Planetary Science Letters, 458, 252, \dodoi{10.1016/j.epsl.2016.10.025}

\bibitem[{{Fung} {et~al.}(2015){Fung}, {Artymowicz}, \& {Wu}}]{Fung2015}
{Fung}, J., {Artymowicz}, P., \& {Wu}, Y. 2015, \apj, 811, 101, \dodoi{10.1088/0004-637X/811/2/101}

\bibitem[{{Genda} \& {Abe}(2005)}]{genda05}
{Genda}, H., \& {Abe}, Y. 2005, \nat, 433, 842, \dodoi{10.1038/nature03360}

\bibitem[{{Haisch} {et~al.}(2001){Haisch}, {Lada}, \& {Lada}}]{haisch01}
{Haisch}, Karl~E., J., {Lada}, E.~A., \& {Lada}, C.~J. 2001, \apjl, 553, L153, \dodoi{10.1086/320685}

\bibitem[{{Hamano} {et~al.}(2013){Hamano}, {Abe}, \& {Genda}}]{hamano13}
{Hamano}, K., {Abe}, Y., \& {Genda}, H. 2013, \nat, 497, 607, \dodoi{10.1038/nature12163}

\bibitem[{{Hayashi}(1981)}]{hayashi81}
{Hayashi}, C. 1981, Progress of Theoretical Physics Supplement, 70, 35, \dodoi{10.1143/PTPS.70.35}

\bibitem[{Hirose {et~al.}(2021)Hirose, Wood, \& Vočadlo}]{Hirose2021}
Hirose, K., Wood, B., \& Vočadlo, L. 2021, Nature Reviews Earth \& Environment, 2, 645, \dodoi{10.1038/s43017-021-00203-6}

\bibitem[{{Iizuka-Oku} {et~al.}(2017){Iizuka-Oku}, {Yagi}, {Gotou}, {Okuchi}, {Hattori}, \& {Sano-Furukawa}}]{iizuka17}
{Iizuka-Oku}, R., {Yagi}, T., {Gotou}, H., {et~al.} 2017, Nature Communications, 8, 14096, \dodoi{10.1038/ncomms14096}

\bibitem[{{Ikoma} \& {Genda}(2006)}]{ikoma06}
{Ikoma}, M., \& {Genda}, H. 2006, \apj, 648, 696, \dodoi{10.1086/505780}

\bibitem[{{Inoue} {et~al.}(1995){Inoue}, {Yurimoto}, \& {Kudoh}}]{Inoue1995}
{Inoue}, T., {Yurimoto}, H., \& {Kudoh}, Y. 1995, \grl, 22, 117, \dodoi{10.1029/94GL02965}

\bibitem[{Ishigaki {et~al.}(2021)Ishigaki, Kominami, Makino, Fujimoto, \& Iwasawa}]{ishigaki21}
Ishigaki, Y., Kominami, J., Makino, J., Fujimoto, M., \& Iwasawa, M. 2021, Publications of the Astronomical Society of Japan, 73, 660, \dodoi{10.1093/pasj/psab028}

\bibitem[{{Iwasawa} {et~al.}(2016){Iwasawa}, {Tanikawa}, {Hosono}, {Nitadori}, {Muranushi}, \& {Makino}}]{iwasawa16}
{Iwasawa}, M., {Tanikawa}, A., {Hosono}, N., {et~al.} 2016, \pasj, 68, 54, \dodoi{10.1093/pasj/psw053}

\bibitem[{{Izidoro} \& {Piani}(2022)}]{Izidoro2022}
{Izidoro}, A., \& {Piani}, L. 2022, Elements, 18, 181, \dodoi{10.2138/gselements.18.3.181}

\bibitem[{{Javoy} {et~al.}(2010){Javoy}, {Kaminski}, {Guyot}, {Andrault}, {Sanloup}, {Moreira}, {Labrosse}, {Jambon}, {Agrinier}, {Davaille}, \& {Jaupart}}]{javoy10}
{Javoy}, M., {Kaminski}, E., {Guyot}, F., {et~al.} 2010, Earth and Planetary Science Letters, 293, 259, \dodoi{10.1016/j.epsl.2010.02.033}

\bibitem[{{Johansen} {et~al.}(2021){Johansen}, {Ronnet}, {Bizzarro}, {Schiller}, {Lambrechts}, {Nordlund}, \& {Lammer}}]{johansen21}
{Johansen}, A., {Ronnet}, T., {Bizzarro}, M., {et~al.} 2021, Science Advances, 7, eabc0444, \dodoi{10.1126/sciadv.abc0444}

\bibitem[{{Kegerreis} {et~al.}(2020){Kegerreis}, {Eke}, {Massey}, \& {Teodoro}}]{Kegerreis2020}
{Kegerreis}, J.~A., {Eke}, V.~R., {Massey}, R.~J., \& {Teodoro}, L.~F.~A. 2020, \apj, 897, 161, \dodoi{10.3847/1538-4357/ab9810}

\bibitem[{{Kokubo} \& {Ida}(2000)}]{kokubo00}
{Kokubo}, E., \& {Ida}, S. 2000, \icarus, 143, 15, \dodoi{10.1006/icar.1999.6237}

\bibitem[{{Kokubo} {et~al.}(2006){Kokubo}, {Kominami}, \& {Ida}}]{kokubo06}
{Kokubo}, E., {Kominami}, J., \& {Ida}, S. 2006, \apj, 642, 1131, \dodoi{10.1086/501448}

\bibitem[{{Kominami} \& {Ida}(2002)}]{kominami02}
{Kominami}, J., \& {Ida}, S. 2002, \icarus, 157, 43, \dodoi{10.1006/icar.2001.6811}

\bibitem[{{Kominami} \& {Ida}(2004)}]{kominami2004}
---. 2004, \icarus, 167, 231, \dodoi{10.1016/j.icarus.2003.10.005}

\bibitem[{Kurokawa {et~al.}(2018)Kurokawa, Foriel, Laneuville, Houser, \& Usui}]{KUROKAWA2018149}
Kurokawa, H., Foriel, J., Laneuville, M., Houser, C., \& Usui, T. 2018, Earth and Planetary Science Letters, 497, 149, \dodoi{https://doi.org/10.1016/j.epsl.2018.06.016}

\bibitem[{{Kurosaki} {et~al.}(2023){Kurosaki}, {Hori}, {Ogihara}, \& {Kunitomo}}]{Kurosaki2023}
{Kurosaki}, K., {Hori}, Y., {Ogihara}, M., \& {Kunitomo}, M. 2023, \apj, 957, 67, \dodoi{10.3847/1538-4357/acfe0a}

\bibitem[{{Kuwahara} {et~al.}(2019){Kuwahara}, {Kurokawa}, \& {Ida}}]{Kuwahara2019}
{Kuwahara}, A., {Kurokawa}, H., \& {Ida}, S. 2019, \aap, 623, A179, \dodoi{10.1051/0004-6361/201833997}

\bibitem[{Kuwayama {et~al.}(2020)Kuwayama, Morard, Nakajima, Hirose, Baron, Kawaguchi, Tsuchiya, Ishikawa, Hirao, \& Ohishi}]{Kuwayama2020}
Kuwayama, Y., Morard, G., Nakajima, Y., {et~al.} 2020, Phys. Rev. Lett., 124, 165701, \dodoi{10.1103/PhysRevLett.124.165701}

\bibitem[{{Lambrechts} \& {Johansen}(2012)}]{lambrechts12}
{Lambrechts}, M., \& {Johansen}, A. 2012, \aap, 544, A32, \dodoi{10.1051/0004-6361/201219127}

\bibitem[{{Matsui} \& {Abe}(1986)}]{matsui86}
{Matsui}, T., \& {Abe}, Y. 1986, \nat, 319, 303, \dodoi{10.1038/319303a0}

\bibitem[{{Matsumoto} {et~al.}(2021){Matsumoto}, {Kokubo}, {Gu}, \& {Kurosaki}}]{Matsumoto2021}
{Matsumoto}, Y., {Kokubo}, E., {Gu}, P.-G., \& {Kurosaki}, K. 2021, \apj, 923, 81, \dodoi{10.3847/1538-4357/ac2b2d}

\bibitem[{{Ormel} {et~al.}(2015){Ormel}, {Shi}, \& {Kuiper}}]{Ormel2015}
{Ormel}, C.~W., {Shi}, J.-M., \& {Kuiper}, R. 2015, \mnras, 447, 3512, \dodoi{10.1093/mnras/stu2704}

\bibitem[{{Ozima} \& {Zahnle}(1993)}]{ozima93}
{Ozima}, M., \& {Zahnle}, K. 1993, Geochemical Journal, 27, 185, \dodoi{10.2343/geochemj.27.185}

\bibitem[{{Pu} \& {Wu}(2015)}]{pu2015}
{Pu}, B., \& {Wu}, Y. 2015, \apj, 807, 44, \dodoi{10.1088/0004-637X/807/1/44}

\bibitem[{{Raymond} {et~al.}(2004){Raymond}, {Quinn}, \& {Lunine}}]{raymond2004}
{Raymond}, S.~N., {Quinn}, T., \& {Lunine}, J.~I. 2004, \icarus, 168, 1, \dodoi{10.1016/j.icarus.2003.11.019}

\bibitem[{{Sakuraba} {et~al.}(2021){Sakuraba}, {Kurokawa}, {Genda}, \& {Ohta}}]{sakuraba2021}
{Sakuraba}, H., {Kurokawa}, H., {Genda}, H., \& {Ohta}, K. 2021, Scientific Reports, 11, 20894, \dodoi{10.1038/s41598-021-99240-w}

\bibitem[{{Tagawa} {et~al.}(2021){Tagawa}, {Sakamoto}, {Hirose}, {Yokoo}, {Hernlund}, {Ohishi}, \& {Yurimoto}}]{Tagawa2021}
{Tagawa}, S., {Sakamoto}, N., {Hirose}, K., {et~al.} 2021, Nature Communications, 12, 2588, \dodoi{10.1038/s41467-021-22035-0}

\bibitem[{{Tanaka} {et~al.}(2002){Tanaka}, {Takeuchi}, \& {Ward}}]{Tanaka2002}
{Tanaka}, H., {Takeuchi}, T., \& {Ward}, W.~R. 2002, \apj, 565, 1257, \dodoi{10.1086/324713}

\bibitem[{Wang {et~al.}(2019)Wang, Zhou, Liu, Sun, Liu, \& Yang}]{Wang2019}
Wang, Y., Zhou, J.-l., Liu, F.-y., {et~al.} 2019, Monthly Notices of the Royal Astronomical Society, 490, 359, \dodoi{10.1093/mnras/stz2375}

\bibitem[{Wu {et~al.}(2019)Wu, Zhang, Zhou, \& Steffen}]{Wu2019}
Wu, D.-H., Zhang, R.~C., Zhou, J.-L., \& Steffen, J.~H. 2019, Monthly Notices of the Royal Astronomical Society, 484, 1538, \dodoi{10.1093/mnras/stz054}

\bibitem[{Yoshinaga {et~al.}(1999)Yoshinaga, Kokubo, \& Makino}]{Yoshinaga1999}
Yoshinaga, K., Kokubo, E., \& Makino, J. 1999, Icarus, 139, 328, \dodoi{https://doi.org/10.1006/icar.1999.6098}

\bibitem[{{Young} {et~al.}(2023){Young}, {Shahar}, \& {Schlichting}}]{young23}
{Young}, E.~D., {Shahar}, A., \& {Schlichting}, H.~E. 2023, \nat, 616, 306, \dodoi{10.1038/s41586-023-05823-0}

\end{thebibliography}
\bibliographystyle{aasjournal}

\end{document}